%% file: main-tr.tex
\documentclass[oneside,letterpaper,10pt]{article}
\pdfoutput=1

\usepackage[letterpaper,total={6.5in, 9in},top=1in]{geometry}

\def\SidewayTables{1}

\input{packages}

\input{macros}

\begin{document}

\input{title}

\author{Venkatesh-Prasad Ranganath \hspace{1cm} Joydeep Mitra\\
Kansas State University, USA\\
\{rvprasad,joydeep\}@k-state.edu
}

\maketitle

\input{abstract}
\sloppy
\input{introduction}
\input{ghera}
\input{ghera-evaluation}
\input{vul-tools-evaluation}
\input{mal-tools-evaluation}
\input{related-work}
\input{artifacts}
\input{future-work}
\input{summary}

\bibliographystyle{plainnat}
\bibliography{references}

\appendix
\input{catalog}

\end{document}

%% file: packages.tex
\usepackage{lmodern}
\usepackage[plain]{fancyref}
\usepackage{xcolor}
\usepackage[hyphens]{url}
\usepackage{hyperref}
\hypersetup{
    colorlinks=true,
    linkcolor=blue,
    filecolor=red,      
    urlcolor=magenta,
    breaklinks=true,
}
\usepackage{paralist}
\usepackage{booktabs}
\usepackage{verbatim}
\usepackage{graphicx}
\usepackage{flushend}
\usepackage{xspace}
\usepackage[figuresright]{rotating}

\usepackage{pifont}
\newcommand{\cmark}{\ding{51}}%

\usepackage{natbib}
\usepackage{amsmath}

%% file: macros.tex
\newcommand{\ie}{i.e.,\xspace}
\newcommand{\eg}{e.g.,\xspace}
\ifundef{\etal}{\newcommand{\etal}{et al.\xspace}}{}

\newcommand{\anonymize}[1]{
  \ifAnonymized {\color{red}{ANONYMIZED}}
  \else #1 
  \fi
}

\newif\ifAnonymized

\newenvironment{customTable}{
  \ifdef{\SidewayTables}{\begin{sidewaystable}}{\begin{table*}}
}{
  \ifdef{\SidewayTables}{\end{sidewaystable}}{\end{table*}}
}

\newcommand*{\fancyrefrqlabelprefix}{rq}
\frefformat{plain}{\fancyrefrqlabelprefix}{RQ#1}
\Frefformat{plain}{\fancyrefrqlabelprefix}{RQ#1}

\newcommand{\amandroid}{Amandroid\xspace}
\newcommand{\androbugs}{AndroBugs\xspace}
\newcommand{\covert}{COVERT\xspace}
\newcommand{\dialdroid}{DIALDroid\xspace}
\newcommand{\devknox}{DevKnox\xspace}
\newcommand{\fixdroid}{FixDroid\xspace}
\newcommand{\flowdroid}{FlowDroid\xspace}
\newcommand{\horndroid}{HornDroid\xspace}
\newcommand{\jaads}{JAADAS\xspace}
\newcommand{\mallodroid}{MalloDroid\xspace}
\newcommand{\marvin}{Marvin-SA\xspace}
\newcommand{\mobsf}{MobSF\xspace}
\newcommand{\qark}{QARK\xspace}
\newcommand{\maldrolyzer}{Maldrolyzer\xspace}
\newcommand{\appcritique}{AppCritique\xspace}
\newcommand{\andrototal}{AndroTotal\xspace}
\newcommand{\hickwall}{HickWall\xspace}
\newcommand{\nviso}{NVISO ApkScan\xspace}
\newcommand{\virustotal}{VirusTotal\xspace}

%% file: title.tex
\title{Are Free Android App Security Analysis Tools Effective in Detecting Known Vulnerabilities?\footnote{This is a pre-print of an article published in \emph{Empirical Software Engineering}. The final authenticated version is available online at: \url{https://doi.org/10.1007/s10664-019-09749-y}.}}

%% file: abstract.tex
\begin{abstract}
Increasing interest in securing the Android ecosystem has spawned numerous efforts to assist app developers in building secure apps.  These efforts have resulted in tools and techniques capable of detecting vulnerabilities (and malicious behaviors) in apps.  However, \emph{there has been no evaluation of the effectiveness of these tools and techniques in detecting known vulnerabilities.}  The absence of such evaluations puts app developers at a disadvantage when choosing security analysis tools to secure their apps.

In this regard, we evaluated the effectiveness of vulnerability detection tools for Android apps.  We reviewed 64 tools and empirically evaluated 14 vulnerability detection tools (incidentally along with five malicious behavior detection tools) against 42 known unique vulnerabilities captured by Ghera benchmarks, which are composed of both vulnerable and secure apps.  Of the 24 observations from the evaluation, the main observation is \emph{existing vulnerability detection tools for Android apps are very limited in their ability to detect known vulnerabilities --- all of the evaluated tools together could only detect 30 of the 42 known unique vulnerabilities}.

More effort is required if security analysis tools are to help developers build secure apps.  We hope the observations from this evaluation will help app developers choose appropriate security analysis tools and persuade tool developers and researchers to identify and address limitations in their tools and techniques.  We also hope this evaluation will catalyze or spark a conversation in the software engineering and security communities to require a more rigorous and explicit evaluation of security analysis tools and techniques.
\end{abstract}

%% file: introduction.tex
\section{Introduction}
\label{sec:introduction}

\subsection{Motivation}
\label{sec:motivation}

Mobile devices have become an integral part of living in present-day society.  They have access to a vast amount of private and sensitive data about their users and, consequently, enable various services for their users such as banking, social networking, and even two-step authentication.  Hence, securing mobile devices and apps that run on them is paramount.

With more than 2 billion Android devices in the world, securing Android devices, platform, and apps is crucial \citep{Sufatrio:2015}.  Companies such as Google and Samsung with access to a large pool of resources are well poised to tackle device and platform security.  However, app stores and app developers share the task of securing Android apps.  While app stores focus on keeping malicious apps out of the ecosystem, malicious apps can enter the ecosystem \eg app installation from untrusted sources, inability to detect malicious behavior in apps, access to malicious websites.  Hence, \emph{there is a need for app developers to secure their apps.}

When developing apps, developers juggle with multiple aspects including app security.  Most app development teams often cannot tend equally well to every aspect as they are often strapped for resources.  Hence, there is an acute need for automatic tools and techniques that can detect vulnerabilities in apps and, when possible, suggest fixes for identified vulnerabilities.

While the software development community has recently realized the importance of security, developer awareness about how security issues transpire and how to avoid them still lacks \citep{Green:IEEESP16}.  Hence, vulnerability detection tools need to be applicable off the shelf with no or minimal configuration.

In this context, numerous efforts have proposed techniques and developed tools to detect different vulnerabilities (and malicious behavior) in Android apps.  Given the number of proposed techniques and available tools, there have been recent efforts to assess the capabilities of these tools and techniques \citep{Reaves:2016,Sadeghi:TSE17,Pauck:FSE18}.  However, these assessments are subject to one or more of the following limiting factors:

\begin{enumerate}
  \item Consider techniques only as reported in the literature, \ie without executing associated tools.
  \item Exercise a small number of tools.
  \item Consider only academic tools.
  \item Consider only tools that employ specific kind of underlying techniques, \eg program analysis.
  \item Rely on technique-specific microbenchmarks, \eg benchmarks targeting the use of taint-flow analysis to detect information leaks.
  \item Rely on benchmarks whose \emph{representativeness} has not been established, \ie \emph{do the benchmarks capture vulnerabilities as they occur in real-world apps?}
  \item Use random real-world apps that are not guaranteed to be vulnerable.
\end{enumerate}

The evaluations performed in efforts that propose new tools and techniques also suffer from such limitations.  Specifically, such evaluations focus on proving the effectiveness of proposed tools and techniques in detecting specific vulnerabilities.  While such a focus is necessary, it is not sufficient as the effectiveness of new tools and techniques in detecting previously known vulnerabilities is unknown.  Hence, the results are limited in their ability to help app developers choose appropriate tools and techniques.

In short, to the best of our knowledge, \emph{there has been no evaluation of the effectiveness of Android app vulnerability detection tools to detect known vulnerabilities without being limited by any of the above factors.}

In addition to the growing dependence on mobile apps, the prevalence of the Android platform, the importance of securing mobile apps, and the need for automatic easy-to-use off-the-shelf tools to build secure mobile apps, here are few more compelling reasons to evaluate the effectiveness of tools in detecting known vulnerabilities in Android apps.

\begin{enumerate}
  \item To develop secure apps, app developers need to choose and use tools that are best equipped to detect the class of vulnerabilities that they believe (based on their intimate knowledge of their apps) will likely plague their apps, \eg based on the APIs used in their apps.  To make good choices, app developers need information about the effectiveness of tools in detecting various classes of vulnerabilities.  Information about other aspects of tools such as performance, usability, and complexity can also be helpful in such decisions.
  \item With the information that a tool cannot detect a specific class of vulnerabilities, app developers can either choose to use a combination of tools to cover all or most of the vulnerabilities of interest or incorporate extra measures in their development process to help weed out vulnerabilities that cannot be detected by the chosen set of tools.
  \item App developers want to detect and prevent known vulnerabilities in their apps as the vulnerabilities, their impact, and their fixes are known a priori.
  \item An evaluation of effectiveness will expose limitations/gaps in the current set of tools and techniques.  This information can aid tool developers to improve their tools.  This information can help researchers direct their efforts to identify the cause of these limitations and either explore ways to address the limitations or explore alternative approaches to prevent similar vulnerabilities, \eg by extending platform capabilities.
\end{enumerate}

In terms of observations from tool evaluations, due to the sixth limiting factor mentioned above, observations from tool evaluations cannot be generalized to real-world apps unless the representativeness of subjects/benchmarks used in tool evaluations has been measured, \ie do the subjects/benchmarks capture vulnerabilities as they occur in real-world apps?  Despite the existence of numerous Android security analysis tools and techniques, \emph{there has been no evaluation of the representativeness of the benchmarks used in evaluating these tools, \ie do the benchmarks capture vulnerabilities as they occur in real-world apps?}

\subsection{Contributions}
\label{sec:contributions}

Motivated by the above observations, \emph{we experimented to evaluate the effectiveness of vulnerability detection tools for Android apps.}  Incidentally, due to the inherent nature of the benchmarks, we also evaluated a few malicious behavior detection tools.  We considered 64 security analysis tools and empirically evaluated 19 of them.  We used benchmarks from Ghera repository \citep{Mitra:PROMISE17} as they captured 42 known vulnerabilities and were known to be tool/technique agnostic, authentic, feature specific, minimal, version specific, comprehensive, and dual \ie contain both vulnerable and malicious apps.

To ensure the observations from the above tools evaluation can be generalized to real-world apps, \emph{we assessed if Ghera benchmarks were representative of real-world apps, \ie do Ghera benchmarks capture vulnerabilities as they occur in real-world apps?}

In this paper, we describe these evaluations, report about the representativeness of Ghera benchmarks, and make 24 observations concerning the effectiveness of Android app security analysis tools.

\subsection{Rest of the Paper}

This paper is organized as follows. \Fref{sec:ghera} describes Ghera repository and our rationale for using it to evaluate tools.  \Fref{sec:ghera-evaluation} describes the experiment to measure the representativeness of Ghera benchmarks along with our observations from the experiment.  \Fref{sec:vul-tools-evaluation} describes the evaluation of the effectiveness of vulnerability detection tools for Android apps along with the 20 observations from the evaluation.  \Fref{sec:mal-tools-evaluation} describes the incidental evaluation of the effectiveness of malicious behavior detection tools for Android apps along with four observations from the evaluation.  \Fref{sec:related-work} discusses prior evaluations of Android security analysis tools and how our evaluation relates to them.  \Fref{sec:artifacts} provides information to access the automation scripts used to perform the evaluation and the artifacts generated in the evaluation.  \Fref{sec:future-work} mentions possible extensions to this effort.  \Fref{sec:summary} summarizes our observations from this evaluation.  \Fref{sec:catalog} briefly catalogs the vulnerabilities captured in Ghera and considered in this evaluation.

%% file: ghera.tex
\section{Ghera: A Repository of Vulnerability Benchmarks}
\label{sec:ghera}

For this evaluation, we considered the Android app vulnerabilities cataloged in \emph{Ghera} repository, a growing repository of benchmarks that captures known vulnerabilities in Android apps \citep{Mitra:PROMISE17}.  We created this repository in 2017 with the goal of cataloging vulnerabilities in Android apps as reported by prior work.

Ghera contains two kinds of benchmarks: \emph{lean} and \emph{fat}.  \emph{Lean benchmarks} are stripped down apps that exhibit vulnerabilities and exploits with almost no other interesting behaviors and functionalities.  \emph{Fat benchmarks} are real-world apps that exhibit specific known vulnerabilities.

At the time of this evaluation, the number of fat benchmarks in Ghera was low.  So, we considered only lean benchmarks in this evaluation.  In the rest of this paper, we will focus on lean benchmarks and refer to them as benchmarks.

Each benchmark capturing a specific vulnerability X is composed of three apps (where applicable): a \emph{benign (vulnerable)} app with vulnerability \emph{X},\footnote{We use the terms benign and vulnerable interchangeably.} a \emph{malicious} app capable of exploiting vulnerability \emph{X} in the benign app, and a \emph{secure} app without vulnerability \emph{X} and, hence, not exploitable by the malicious app.  Malicious apps are absent in benchmarks when malicious behavior occurs outside the Android environment, \eg web app.
Each benchmark is accompanied by instructions to demonstrate the captured vulnerability and the corresponding exploit by building and executing the associated apps.  Consequently, the presence and absence of vulnerabilities and exploits in these benchmarks are verifiable.

At the time of this evaluation, Ghera contained 42 benchmarks grouped into the following seven categories based on the nature of the APIs (including features of XML-based configuration) involved in the creation of captured vulnerabilities.  (Category labels that are used later in the paper appear in square brackets.)

\begin{enumerate}
  \item \emph{Crypto} category contains 4 vulnerabilities involving cryptography API. [Crypto]
  \item \emph{ICC} category contains 16 vulnerabilities involving inter-component communication (ICC) API. [ICC]
  \item \emph{Networking} category contains 2 vulnerabilities involving networking (non-web) API. [Net]
  \item \emph{Permission} category contains 1 vulnerability involving permission API. [Perm]
  \item \emph{Storage} category contains 6 vulnerabilities involving data storage and SQL database APIs. [Store]
  \item \emph{System} category contains 4 vulnerabilities involving system API dealing with processes. [Sys]
  \item \emph{Web} category contains 9 vulnerabilities involving web API. [Web]
\end{enumerate}

\Fref{sec:catalog} briefly catalogs these vulnerabilities and their canonical references.

\subsection{Why Use Ghera?}
\label{sec:ghera-why-use}

Past efforts focused on the creation of benchmarks have considered criteria to ensure/justify the benchmarks are useful, \eg database related benchmarks have considered relevance, scalability, portability, ease of use, ease of interpretation, functional coverage, and selectivity coverage \citep{Gray:Book92}; web services related benchmarks have considered criteria such as repeatability, portability, representativeness, non-intrusiveness, and simplicity \citep{Antunes:ICWS10}.

In a similar spirit, for evaluations of security analysis tools to be useful to tool users, tool developers, and researchers, we believe evaluations should be based on vulnerabilities (consequently, benchmarks) that are \emph{valid} (\ie will result in a weakness in an app), \emph{general} (\ie do not depend on uncommon constraints such as rooted device or admin access), \emph{exploitable} (\ie can be used to inflict harm), and \emph{current} (\ie occur in existing apps and can occur in new apps).

The vulnerabilities captured in Ghera benchmarks have been previously reported in the literature or documented in Android documentation; hence, they are \emph{valid}.  These vulnerabilities can be verified by executing the benign and malicious apps in Ghera benchmarks on vanilla Android devices and emulators; hence, they are \emph{general} and \emph{exploitable}.  These vulnerabilities are \emph{current} as they are based on Android API levels 19 thru 25, which enable more than $90\%$ of Android devices in the world and are targeted by both existing and new apps.

Due to these characteristics and the salient characteristics of Ghera --- \emph{tool and technique agnostic, authentic, feature specific, contextual (lean), version specific, duality} and \emph{comprehensive} --- described by \citet*{Mitra:PROMISE17}, Ghera is well-suited for this evaluation.

%% file: ghera-evaluation.tex
\section{Representativeness of Ghera Benchmarks}
\label{sec:ghera-evaluation}

For tools evaluations to be useful, along with the above requirements, \emph{the manifestation of a vulnerability considered in the evaluation should be representative of the manifestations of the vulnerability in real-world apps.}  In other words, \emph{the evaluation should consider vulnerabilities as they occur in real-world apps.}

A vulnerability can \emph{manifest or occur} in different ways in apps due to various aspects such as producers and consumers of data, nature of data, APIs involved in handling and processing data, control/data flow paths connecting various code fragments involved in the vulnerability, and platform features involved in the vulnerability.  As a simple example, consider a vulnerability that leads to information leak: sensitive data is written into an insecure location.  This vulnerability can manifest in multiple ways.  Specifically, at the least, each combination of different ways of writing data into a location (\eg using different I/O APIs) and different insecure locations (\eg insecure file, untrusted socket) can lead to \emph{a unique manifestation of the vulnerability}.

In terms of representativeness as described above and as desired in tool assessments (as mentioned in \Fref{sec:motivation}), there is no evidence that the benchmarks in Ghera capture vulnerabilities as they manifest or occur in real-world apps; hence, we needed to establish the representativeness of Ghera benchmarks.

\subsection{How to Measure Representativeness?}
\label{sec:ghera-how-to-representativeness}

Since Ghera benchmarks capture specific manifestations of known vulnerabilities, we wanted to identify these manifestations in real-world apps to establish the representativeness of the benchmarks.  However, there was no definitive list of versions of apps that exhibit known vulnerabilities.  So, we explored CVE \citep{CVE:URL}, an open database of vulnerabilities discovered in real-world Android apps, to identify vulnerable versions of apps.  We found that most CVE vulnerability reports failed to provide sufficient information about the validity, exploit-ability, and manifestation of vulnerabilities in the reported apps.  Next, we considered the option of manually examining apps mentioned in CVE reports for vulnerabilities.  This option was not viable because CVE vulnerability reports do not include copies of reported apps.  Also, while app version information from CVE could be used to download apps for manual examination, only the latest version of apps are available from most Android app stores and app vendors.

Finally, we decided to use usage information about Android APIs involved in manifestations of vulnerabilities as a proxy to establish the representativeness of Ghera benchmarks.  The rationale for this decision is the likelihood of a vulnerability occurring in real-world apps is directly proportional to the number of real-world apps using the Android APIs involved in the vulnerability.  So, \emph{as a weak yet general measure of representativeness, we identified Android APIs used in Ghera benchmarks and measured how often these APIs were used in real-world apps}.

\subsection{Experiment}
\label{sec:ghera-experiment}

\subsubsection{Source of Real-World Apps} We used AndroZoo as the source of real-world Android apps.  AndroZoo is a growing collection of Android apps gathered from several sources including the official Google Play store \citep{Allix:MSR16}.  In May 2018, AndroZoo contained more than 5.8 million different APKs (app bundles).

Every APK (app bundle) contains an XML-based manifest file and a DEX file that contains the code and data (\ie resources, assets)  corresponding to the app.  By design, each Android app is self-contained.  So, the DEX file contains all code that is necessary to execute the app but not provided by the underlying Android Framework or Runtime.  Often, DEX files include code for Android support library.

AndroZoo maintains a list of all of the gathered APKs.  This list documents various features of APKs such as SHA256 hash of an APK (required to download the APK from AndroZoo), size of an APK, and the date (\emph{dex\_date}) associated with the contained DEX file.\footnote{\emph{dex\_date} may not correspond to the release date of the app.}  However, this list does not contain information about API levels (Android versions) that are targeted by the APKs; this information can be recovered from the APKs.

\subsubsection{App Sampling} Each version of Android is associated with an API level, \eg Android versions 5.0 and 5.1 are associated with API levels 21 and 22, respectively.  Every Android app is associated with a minimum API level (version) of Android required to use the app and a target API level (version) of Android that is ideal to use the app; this information is available in the app's manifest file.

At the time of this tools evaluation, Ghera benchmarks targeted API levels 19 thru 25 excluding 20.\footnote{API level 20 was excluded because it is API level 19 with wearable extensions and Ghera benchmarks focus on vulnerabilities in Android apps running on mobile devices.}\footnote{In the rest of this manuscript, ``API levels 19 thru 25'' means API levels 19 thru 25 excluding 20.}  So, we decided to select only apps that targeted API level 19 or higher and required minimum API level 14 or higher.\footnote{We chose API level 14 as the cut-off for minimum API level as the number of apps targeting API level 19 peaked at minimum API level 14.}  Since minimum and target API levels of apps were not available in the AndroZoo APK list, we decided to select apps based on their release dates.  As API level 19 was released in November 2014, we decided to select only apps that were released after 2014.  Since release dates of APKs were not available from AndroZoo APK list, we decided to use dex\_date as a proxy for the release date of apps.

Based on the above decisions, we analyzed the list of APKs available from AndroZoo to select the APKs to download.  We found 2.3 million APKs with dex\_date between 2015 and 2018 (both inclusive).  In these APKs, there were 790K, 1346K, 156K, and 17K APKs with dex\_date from the years 2015, 2016, 2017, and 2018, respectively.  From these APKs, we drew an unequal probability sample without replacement and with probabilities 0.1, 0.2, 0.4, and 0.8 of selecting an APK from years 2015 thru 2018, respectively.  We used \emph{unequal probability sampling} to give preference to the latest APKs as the selected apps would likely target recent API levels and to adjust for the non-uniformity of APK distribution across years.  To create a sample with at least 100K real-world Android apps that targeted the chosen API levels, we tried to download 339K APKs and ended up downloading 292K unique APKs.  Finally, we used the \texttt{apkanalyzer} tool from Android Studio to identify and discard downloaded apps (APKs) with target API level less than 19 or minimum API level less than 14.  \emph{This resulted in a sample of 111K real-world APKs that targeted API levels 19 thru 25 (excluding 20) or higher.}

\subsubsection{API-based App Profiling} Android apps access various capabilities of the Android platform via features of XML-based manifest files and Android programming APIs.  \emph{We refer to the published Android programming APIs and the XML elements and attributes (features) of manifest files collectively as APIs.}  We use the term API to mean either a function, a method, a field, or an XML feature.

For each app (APK), we collected its API profile based on the APIs that were used by or defined in the app, and we deemed as \emph{relevant} to this evaluation as follows.

\begin{enumerate}
  \item From the list of elements and attributes that can be present in a manifest file, based on our knowledge of Ghera benchmarks, we conservatively identified the values of 7 attributes (\eg \textit{intent-filter/category@name}), the presence of 26 attributes (\eg \textit{path-permission@\linebreak[0]writePermission}), and the presence of 6 elements (\eg \textit{uses-permission}) as APIs relevant to this evaluation.  For each app, we recorded the APIs used in the app's manifest.
  \item For an app, we considered all published (\ie public and protected) methods along with all methods that were used but not defined in the app.  Former methods accounted for callback APIs provided by an app and latter methods accounted for service offering (external) APIs used by an app.  We also considered all fields used in the app.  From these considered APIs, we discarded obfuscated APIs, \ie with a single character name.  To make apps comparable in the presence of definitions and uses of overridden Java methods (APIs), if a method was overridden, then we considered the fully qualified name (FQN) of the overridden method in place of the FQN of the overriding method using Class Hierarchy Analysis.  Since we wanted to measure representativeness in terms of Android APIs, we discarded APIs whose FQN did not have any of these prefixes: \textit{java, org, android,} and \textit{com.android}.  For each app, we recorded the remaining APIs.
  \item Numerous APIs commonly used in almost all Android apps are related to aspects (\eg UI rendering) that are not the focus of Ghera benchmarks.  To avoid their influence on the result, we decided to ignore such APIs.  So, we considered the benign app of the template benchmark in Ghera repository.  This app is a basic Android app with one activity containing a couple of widgets and no functionality.  Out of the 1502 APIs used in this app, we manually identified 1134 APIs as commonly used in Android apps (almost all of them were basic Java APIs or related to UI rendering and XML processing).  For each app, we removed these APIs from its list of APIs recorded in above steps 1 and 2 and considered the remaining APIs as relevant APIs.
\end{enumerate}

To collect API profiles of apps in Ghera, we used the APKs available in Ghera repository because we had eliminated extraneous APIs from these APKs by using the \texttt{proguard} tool.

While collecting API-based profile of apps in AndroZoo sample, we discarded 5\% of the APKs due to errors in APKs (\eg missing required attributes) and tooling issues (\eg parsing errors).  \emph{Finally, we ended up with a sample of 105K real-world APKs (apps) from AndroZoo.}

\subsubsection{Measuring Representativeness} We identified the set of relevant APIs associated with benign apps in Ghera using the steps described in the previous section.  \emph{Of the resulting 601 unique relevant APIs, we identified 117 as security-related APIs.}  Based on our experience with Ghera, we were certain the identified APIs could influence app security.  For both these sets of APIs, we measured representativeness in two ways.

\begin{enumerate}
  \item \textit{Using API Use Percentage.} For each API, we calculated the percentage of sample apps that used it.

    To observe how representativeness changes across API levels, we created API level specific app samples.  The app sample specific to API level $k$ contained every sample app whose minimum API level was less than or equal to $k$, and the target API level was greater than or equal to $k$.  In each API level specific sample, for each API, we calculated the percentage of apps that used the API.

    The rationale for this measurement is, \emph{if Ghera benchmarks are representative of real-world apps in terms of using an API, then a large number of real-world apps should use the API.}

  \item \textit{Comparing Sampling Proportions.} For each API, we calculated the sampling proportion of sample apps that used the API.

    To calculate the sampling proportion, we randomly selected 80\% of the sample apps, grouped them into sub-samples containing 40 apps each, and calculated the mean of the proportions in each sub-sample.  We also calculated the sampling proportion of benign apps in Ghera that used the API by randomly drawing 40 samples (with replacement) containing 40 apps each.  We then compared the sampling proportions with confidence level = 0.95, p-value $\leq$ 0.05, and the null hypothesis being the proportion of benign apps in Ghera using the API is less than or equal to the proportion of real-world apps using the API.

  We performed this test both at specific API levels and across API levels.

  The rationale for this measurement is, \emph{if Ghera benchmarks are representative of real-world apps in terms of using an API, then the proportion of Ghera benchmarks using the API should be less than or equal to the proportion of real-world apps using the API.}
\end{enumerate}

\textit{With Top-200 Apps.} We gathered the top-200 apps from Google Play store on April 18, 2018, and repeated the above measurements both across API levels and at specific API levels.  Only 163 of the top-200 apps made it thru the app sampling and API-based app profiling process due to API level restrictions, errors in APKs, and tooling issues.  Hence, we considered 40 sub-samples with replacement containing 40 apps each to measure representativeness by comparing sampling proportions.

\subsection{Observations}
\label{sec:ghera-observations}

\subsubsection{Based on API Use Percentage}  
\begin{figure*}
  \centering
  \includegraphics[width=\textwidth]{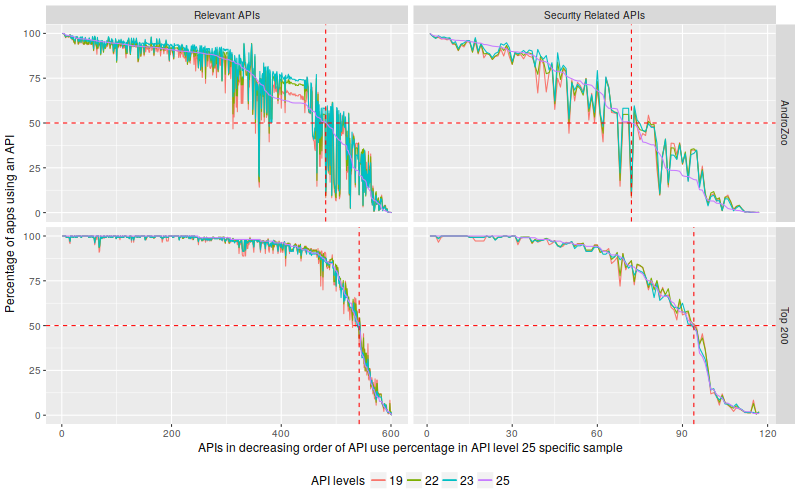}
  \caption{Percentage of apps that use APIs used in benign Ghera apps.}
  \label{fig:representativeness}
\end{figure*}

The color graphs in \Fref{fig:representativeness} show the percentage of sample real-world Android apps using the APIs used in benign apps in Ghera.  The Y-axis denotes the percentage of apps using an API.  The X-axis denotes APIs in decreasing order of percentage of their usage in API level 25 specific sample.  The graphs on the left are based on relevant APIs while the graphs on the right are based only on security-related APIs.  The graphs on the top are based on the AndroZoo sample while the graphs on the bottom are based on the top-200 sample.  To avoid clutter, we have not plotted data for API levels 21 and 24 as they were closely related to API levels 22 and 25, respectively.

Since API level 25 is the latest API level considered in this evaluation, we focus on the data for API level 25 specific samples.  In AndroZoo sample, we observe 80\% (481 out of 601) of relevant APIs used in benign apps in Ghera were each used by more than 50\% (52K) of real-world apps; see the dashed red lines in the top left graph in \Fref{fig:representativeness}.  For the top-200 sample, this number increases to 90\% (542 out of 601); see the dashed red lines in the bottom left graph in \Fref{fig:representativeness}.  When considering only security-related APIs, 61\% (72 out of 117) of APIs used in benign apps in Ghera were each used by more than 50\% of real-world apps in AndroZoo sample; see the dashed red lines in the top right graph in \Fref{fig:representativeness}.  For the top-200 sample, this number increases to 80\% (94 out of 117); see the dashed red lines in the bottom right graph in \Fref{fig:representativeness}.

Barring few APIs in case of AndroZoo sample (\ie dips for APIs ranked between 300 and 400 and closer to 481 (vertical dotted line) in the top left graph in \Fref{fig:representativeness}), the above observations hold true for all API levels considered from 19 thru 25 in both the AndroZoo sample and the top-200 sample.

Further, while we do not present the data in this manuscript, we made similar observations in cases of malicious apps and secure apps in Ghera with both the AndroZoo sample and the top-200 sample.\footnote{The raw data supporting these observations is available as part of publicly available evaluation artifacts; see \Fref{sec:artifacts}.}

Above observations suggest \emph{a large number of real-world apps use a large number of APIs used in Ghera benchmarks.  Consequently, we can conclude that Ghera benchmarks are representative of real-world apps}.

\paragraph{An Opportunity with Security-Related APIs} In the bottom right graph in \Fref{fig:representativeness}, observe that a large percentage of top-200 apps use many security-related APIs.  This large percentage is no surprise as these apps are likely to be widely used and are likely built with proper security measures to keep their users secure.  Surprisingly, based on the AndroZoo sample (see top right graph in \Fref{fig:representativeness}), this is also true of real-world apps --- 72 of the 117 security-related APIs are each used by more than 52K apps from AndroZoo sample in all considered API levels.  Thus, \emph{a large number of real-world apps use security-related APIs knowingly or unknowingly and correctly or incorrectly}.  Hence, there is a tremendous opportunity to help identify and fix incorrect use of security-related APIs.

\subsubsection{Based on Sampling Proportions} Columns 4, 5, 8, and 9 in \Fref{tab:ghera-api-repr} report the number of relevant APIs for which the null hypothesis --- the proportion of benign apps in Ghera using API X is less than or equal to the proportion of real-world apps using API X --- could not be rejected.  This data suggests, for at least 68\% of relevant APIs, the proportion of Ghera apps that used an API could be less than or equal to the proportion of real-world apps in the AndroZoo sample that use the same API.  This is true across all API levels and at specific API levels.  This is also true for at least 94\% of security-related APIs.  In the case of the top-200 sample, this is true for at least 94\% of relevant APIs and 99\% of security-related APIs.\footnote{These numbers will only be higher when lower p-value thresholds are considered.}

\begin{table*}
  \centering
  \ifdef{\TopCaption}{
    \caption{Representativeness based on sample proportions test.  Of the 601 selected (all) APIs, 117 APIs were security-related.}
  }{}
  \begin{tabular}{@{}crrccrrcc@{}}
    \toprule
    & \multicolumn{4}{c}{AndroZoo Apps} & \multicolumn{4}{c}{top-200 Apps} \\
    \cmidrule(r){2-5} \cmidrule(r){6-9}
    & & & \multicolumn{2}{c}{Representative} & & & \multicolumn{2}{c}{Representative} \\
    & & & \multicolumn{2}{c}{APIs (\%)} & & & \multicolumn{2}{c}{APIs (\%)} \\
    \cmidrule(r){4-5} \cmidrule(r){8-9}
    API & Sample & Sub- & Relevant & Security & Sample & Sub- & Relevant & Security \\
    Level & Size & Samples & & Related & Size & Samples & & Related \\
    \midrule
  19-25 & 105586 & 2111 & 414 (68) & 110 (94) & 163 & 40 & 571 (95) & 116 (99) \\
    \midrule
  19 & 105483 & 2109 & 415 (69) & 110 (94) & 143 & 40 & 569 (94) & 116 (99) \\
  21 & 93673 & 1873 & 438 (72) & 110 (94) & 156 & 40 & 565 (94) & 116 (99) \\
  22 & 78470 & 1569 & 450 (74) & 111 (94) & 153 & 40 & 569 (94) & 116 (99) \\
  23 & 55087 & 1101 & 464 (77) & 113 (96) & 128 & 40 & 567 (94) & 116 (99) \\
  24 & 7530 & 150 & 466 (77) & 113 (96) & 109 & 40 & 568 (94) & 116 (99) \\
  25 & 2863 & 57 & 430 (71) & 112 (95) & 104 & 40 & 568 (94) & 116 (99) \\
    \bottomrule
  \end{tabular}
  \ifundef{\TopCaption}{
    \caption{Representativeness based on sample proportions test.  Of the 601 selected (all) APIs, 117 APIs were security-related.}
  }{}
  \label{tab:ghera-api-repr}
\end{table*}

Further, we made similar observations in cases of malicious apps and secure apps in Ghera with both the AndroZoo sample and the top-200 sample.

\emph{Considering Ghera benchmarks as a custom sample in which the proportion of benchmarks that used a specific set of APIs (relevant or security-related) was expected to be high, the above observations suggest such proportions are higher for these APIs in real-world apps.  Consequently, we can conclude that Ghera benchmarks are representative of real-world apps.}

\subsection{Threats to Validity}
\label{sec:ghera-threat-to-validity}

This evaluation of representativeness is based on a weak measure of manifestation of vulnerabilities --- the use of APIs involved in vulnerabilities.  Hence, this evaluation could have ignored the influence of richer aspects such as API usage context, security desiderata of data, and data/control flow path connecting various API uses, which are involved and hard to measure.   The influence of such aspects can be verified by measuring representativeness by considering these aspects.

While we considered a large sample of real-world apps, the distribution of apps across targeted API levels was skewed --- there were fewer apps targeting recent API levels.  Hence, recent API level specific samples may not have exhibited the variations observed in larger API specific samples, \eg API level 19 (see \Fref{fig:representativeness}).  This possibility can be verified by repeating this experiment using samples of comparable sizes.

The version of Ghera benchmarks considered in this evaluation was developed when API level 25 was the latest Android API level.  So, it is possible that tooling and library support available for API level 25 could have influenced the structure of Ghera benchmarks and, consequently, the observations in \Fref{sec:ghera-experiment}.  This possibility can be verified by repeating this experiment in the future while using tooling and library support available with newer API levels.

We have taken extreme care to avoid errors while collecting and analyzing data.  Even so, there could be errors in our evaluation in the form of commission errors (\eg misinterpretation of data, misidentification of APIs as relevant and security-related), omission errors (\eg missed data points), and automation errors.  This threat to validity can be addressed by examining both our raw and processed data sets (see \Fref{sec:artifacts}), analyzing the automation scripts, and repeating the experiment.

%% file: vul-tools-evaluation.tex
\section{Vulnerability Detection Tools Evaluation}
\label{sec:vul-tools-evaluation}

\subsection{Android App Security Analysis Solutions}
\label{sec:vul-tools-solutions}

\anonymize{While creating Ghera repository in 2017, we became aware of various solutions for Android app security.}  From June 2017, we started collecting information about such solutions.  Our primary sources of information were research papers \citep{Sadeghi:TSE17,Reaves:2016}, repositories \citep{AshishGitHub:URL}, and blog posts \citep{MobileWiki:URL} that collated information about Android security solutions.

From these sources, we reviewed 64 solutions that were related to Android security.  (The complete list of these solutions is available at \url{https://bitbucket.org/secure-it-i/may2018}.)  We classified the solution along the following dimensions.

\begin{enumerate}
  \item \textit{Tools vs Frameworks:} \emph{Tools} detect a fixed set of security issues.  While they can be applied immediately, they are limited to detecting a fixed set of issues.  On the other hand, \emph{frameworks} facilitate the creation of tools that can detect specific security issues.  While they are not immediately applicable to detect vulnerabilities and require effort to create tools, they enable detection of a relatively open set of issues.

  \item \textit{Free vs. Commercial:} Solutions are available either freely or for a fee.

  \item \textit{Maintained vs. Unmaintained:} Solutions are either actively maintained or unmaintained (\ie few years of development dormancy).  Typically, unmaintained solutions do not support currently supported versions of Android. This observation is also true of a few maintained solutions.

  \item \textit{Vulnerability Detection vs. Malicious Behavior Detection:} Solutions either detect vulnerabilities in an app or flag signs of malicious behavior in an app.  App developers typically use the former while app stores and end users use the latter.

  \item \textit{Static Analysis vs. Dynamic Analysis:} Solutions that rely on \emph{static analysis} analyze either source code or Dex bytecode of an app and provide verdicts about possible security issues in the app.  Since static analysis abstracts the execution environment, program semantics, and interactions with users and other apps, solutions reliant on static analysis can detect issues that occur in a variety of settings.  However, since static analysis may incorrectly consider invalid settings due to too permissive abstractions, they are prone to high false positive rate.

    In contrast, solutions that rely on \emph{dynamic analysis} execute apps and check for security issues at runtime.  Consequently, they have a very low false positive rate.  However, they are often prone to a high false negative rate because they may fail to explore specific settings required to trigger security issues in an app.

  \item \textit{Local vs. Remote:} Solutions are available as executables or as sources from which executables can be built.  These solutions are installed and executed locally by app developers.  

    Solutions are also available remotely as web services (or via web portals) maintained by solution developers.  App developers use these services by submitting the APKs of their apps for analysis and later accessing the analysis reports.  Unlike local solutions, app developers are not privy to what happens to their apps when using remote solutions.

\end{enumerate}

\subsection{Tools Selection}
\label{sec:vul-tools-selection}

\subsubsection{Shallow Selection}
\label{sec:vul-shallow-selection}

To select tools for this evaluation, we first screened the considered 64 solutions by reading their documentation and any available resources.  We rejected five solutions because they lacked proper documentation, \eg no documentation, lack of instructions to build and use tools.  This elimination was necessary to eliminate human bias resulting from the effort involved in discovering how to build and use a solution, \eg DroidLegacy \citep{Deshotels:PPREQ12}, BlueSeal \citep{Shen:ASE14}.  We rejected AppGuard \citep{AppGuard:URL} because its documentation was not in English.  We rejected six solutions such as Aquifer \citep{Nadkarni:USENIX16}, Aurasium \citep{Xu:USENIX12}, and FlaskDroid \citep{Buguel:USENIX13} as they were not intended to detect vulnerabilities, \eg intended to enforce security policy.

Of the remaining 52 solutions, we selected solutions based on the first four dimensions mentioned in \Fref{sec:vul-tools-solutions}.

In terms of tools vs. frameworks, we were interested in solutions that could readily detect vulnerabilities with minimal adaptation, \ie use it off-the-shelf instead of having to build an extension to detect a specific vulnerability.  The rationale was to eliminate human bias and errors involved in identifying, creating, and using the appropriate adaptations.  Also, we wanted to mimic a \emph{simple developer workflow}: procure/build the tool based on APIs it tackles and the APIs used in an app, follow its documentation, and apply it to the app.  Consequently, we rejected 16 tools that only enabled security analysis, \eg Drozer \citep{Drozer:URL}, ConDroid \citep{Anand:2012}.  When a framework provided pre-packaged extensions to detect vulnerabilities, we selected such frameworks and treated each such extension as a distinct tool, \eg we selected \amandroid framework as it comes with seven pre-packaged vulnerability detection extensions (\ie data leakage ($\amandroid_1$), intent injection ($\amandroid_2$), comm leakage ($\amandroid_3$), password tracking ($\amandroid_4$), OAuth tracking ($\amandroid_5$), SSL misuse ($\amandroid_6$), and crypto misuse \linebreak ($\amandroid_7$)) that can each be used as a separate off-the-shelf tool ($\amandroid_x$) \citep{Wei:2014}.

In terms of free vs. commercial, we rejected AppRay as it was a commercial solution \citep{AppRay:URL}.  While \appcritique was a commercial solution, a feature-limited version of it was available for free.  We decided to evaluate the free version and did not reject \appcritique.

In terms of maintained vs. unmaintained, we focused on selecting only maintained tools.  So, we rejected AndroWarn and ScanDroid tools as they were not updated after 2013 \citep{Androwarn:URL,Fuchs:09}.  In a similar vein, since our focus was on currently supported Android API levels, we rejected TaintDroid as it supported only API levels 18 or below \citep{Enck:2010}.

In terms of vulnerability detection and malicious behavior detection, we selected only vulnerability detection tools and rejected six malicious behavior detection tools.

Next, we focused on tools that could be applied as is without extensive configuration (or inputs).  The rationale was to eliminate human bias and errors involved in identifying and using appropriate configurations.  So, we rejected tools that required additional inputs to detect vulnerabilities.  Specifically, we rejected Sparta as it required analyzed apps to be annotated with security types \citep{Ernst:CCS14}.

Finally, we focused on the applicability of tools to Ghera benchmarks.\footnote{In the rest of this manuscript, for brevity, we say \emph{``a tool is applicable to a benchmark''} if the tool is designed or claims to detect the vulnerability captured by the benchmark.  Likewise, \emph{``a tool is applicable to a set of benchmarks''} if the tool is applicable to at least one benchmark in the set.}  We considered only tools that claimed to detect vulnerabilities stemming from APIs covered by at least one category in Ghera benchmarks.  For such tools, based on our knowledge of Ghera benchmarks and shallow exploration of the tools, we assessed if the tools were indeed applicable to the benchmarks in the covered categories.  The shallow exploration included checking if the APIs used in Ghera benchmarks were mentioned in any lists of APIs bundled with tools, \eg the list of the information source and sink APIs bundled with \horndroid and \flowdroid.  In this regard, we rejected 2 tools (and 1 extension): a) PScout \citep{Au:2012} focused on vulnerabilities related to over/under use of permissions and the only permission related benchmark in Ghera was not related to over/under use of permissions and b) LetterBomb \citep{Garcia:2017} and \amandroid's OAuth tracking extension ($\amandroid_5$) as they were not applicable to any Ghera benchmark.\footnote{While Ghera did not have benchmarks that were applicable to some of the rejected tools at the time of this evaluation, it currently has such benchmarks that can be used in subsequent iterations of this evaluation.}

\subsubsection{Deep Selection}
\label{sec:vul-deep-selection}

Of the remaining 23 tools, for tools that could be executed locally, we downloaded the latest official release of the tools, \eg \amandroid.

If an official release was not available, then we downloaded the most recent version of the tool (executable or source code) from the master branch of its repository, \eg \androbugs.  We then followed the tool's documentation to build and set up the tool.  If we encountered issues during this phase, then we tried to fix them; specifically, when issues were caused by dependency on older versions of other tools (\eg \horndroid failed against real-world apps as it was using an older version of \texttt{apktool}, a decompiler for Android apps), incorrect documentation (\eg documented path to the \dialdroid executable was incorrect), and incomplete documentation (\eg IccTA's documentation did not mention the versions of required dependencies \citep{Li:2015}).  Our fixes were limited to being able to execute the tools and not to affect the behavior of the tool.  We stopped trying to fix an issue and rejected a tool if we could not figure out a fix by interpreting the error messages and by exploring existing publicly available bug reports.  This policy resulted in rejecting DidFail \citep{Klieber:SOAP14}.

Of the remaining tools, we tested 18 local tools using the benign apps from randomly selected Ghera benchmarks I1, I2, W8, and W9\footnote{Refer to \Fref{sec:catalog} for a brief description of these benchmarks.} and randomly selected apps Offer Up, Instagram, Microsoft Outlook, and My Fitness Pal's Calorie Counter from Google Play store.  We execute each tool with each of the above apps as input on a 16 core Linux machine with 64GB RAM and with 15 minute time out period.  If a tool failed to execute successfully on all of these apps, then we rejected the tool. Explicitly, we rejected IccTA and SMV Hunter because they failed to process the test apps by throwing exceptions \citep{Li:2015,Sounthiraraj:2014}.  We rejected CuckooDroid and DroidSafe because they ran out of time or memory while processing the test apps \citep{Gordon:2015,CuckooDroid:URL}.

For nine tools that were available only remotely, we tested them by submitting the above test apps for analysis. If a tool's web service was unavailable, failed to process all of the test apps, or did not provide feedback within 30--60 minutes, then we rejected it.  Consequently, we rejected four remote tools, \eg TraceDroid \citep{Tracedroid:URL}.

\subsubsection{Selected Tools} \Fref{tab:vul-tools-info} lists the fourteen tools selected for evaluation along with their canonical reference.  For each tool, the table reports the version (or the commit id) selected for evaluation, dates of its initial publication and latest update, whether it uses static analysis (S) or dynamic analysis (D) or both (SD), whether it runs locally (L) or remotely (R), whether it is an academic (A) or non-academic (N) tool, whether it uses shallow analysis (H) or deep analysis (E), and the time spent to set up tools on a Linux machine.

\begin{customTable}
  \centering
  \ifdef{\TopCaption}{
    \caption{Evaluated vulnerability detection tools.  ``?'' denotes unknown information.  \emph{S} and \emph{D} denote use of static analysis and dynamic analysis, respectively.  \emph{L} and \emph{R} denote the tool runs locally and remotely, respectively.  \emph{A} and \emph{N} denote academic tool and non-academic tool, respectively.  \emph{H} and \emph{E} denote the tool uses shallow analysis and deep analysis, respectively.  Empty cell denotes non-applicable cases.}
  }{}
  \begin{tabular}{@{}lccccccr@{}}
    \toprule
    Tool & Commit Id / & Updated & S/D & L/R & A/N & H/E & Set Up\\
    & Version & [Published] & & & & & Time (sec)\\
    \midrule
    \amandroid \citep{Wei:2014} & 3.1.2 & 2017 [2014] & S & L & A & E & 3600 \\

    \androbugs \citep{AndroBugs:URL} & 7fd3a2c & 2015 [2015] & S & L & N & H & 600 \\

    \appcritique \citep{AppCritique:URL} & ? & ? [?] & ? & R & N & ? & \\

    \covert \citep{Bagheri:TSE2015} & 2.3 & 2015 [2015] & S & L & A & E & 2700\\

    \devknox \citep{Devknox:URL} & 2.4.0 & 2017 [2016] & S & L & N & H & 600 \\

    \dialdroid \citep{Bosu:2017} & 25daa37 & 2018 [2016] & S & L & A & E & 3600 \\

    \fixdroid \citep{Nguyen:2017} & 1.2.1 & 2017 [2017] & S & L & A & H & 600 \\

    \flowdroid \citep{Arzt:2014} & 2.5.1 & 2018 [2013] & S & L & A & E & 9000 \\

    \horndroid \citep{Calzavara:2017} & aa92e46 & 2018 [2017] & S & L & A & E & 600 \\

    \jaads \citep{Jaads:URL} & 0.1 & 2017 [2017] & S & L & N & H/E & 900\\

    \mallodroid \citep{Fahl:2012} & 78f4e52 & 2013 [2012] & S & L & A & H & 600 \\

    \marvin\footnote{We refer to Marvin Static Analyzer as \marvin.} \citep*{Marvin:URL} & 6498add & 2016 [2016] & S & L & N & H & 600 \\

    \mobsf \citep{MobSF:URL} & b0efdc5 & 2018 [2015] & SD & L & N & H & 1200 \\

    \qark \citep{Qark:URL} & 1dd2fea & 2017 [2015] & S & L & N & H & 600\\
    \bottomrule
  \end{tabular}
  \ifundef{\TopCaption}{
    \caption{Evaluated vulnerability detection tools.  ``?'' denotes unknown information.  \emph{S} and \emph{D} denote use of static analysis and dynamic analysis, respectively.  \emph{L} and \emph{R} denote the tool runs locally and remotely, respectively.  \emph{A} and \emph{N} denote academic tool and non-academic tool, respectively.  \emph{H} and \emph{E} denote the tool uses shallow analysis and deep analysis, respectively.  Empty cell denotes non-applicable cases.}
  }{}
  \label{tab:vul-tools-info}
\end{customTable}

\subsection{Experiment}
\label{sec:vul-tools-experiment}

Every selected vulnerability detection tool (including pre-packaged extensions treated as tools) was applied in its default configuration to the \textit{benign} app and the \textit{secure} app separately of every applicable Ghera benchmark (given in column 9 in \Fref{tab:vul-evals}).
To control for the influence of API level on the performance of tools, we used APKs that were built with minimum API level of 19 and target API level of 25, \ie these APKs can be installed and executed with every API level from 19 thru 25.

We executed the tools on a 16 core Linux machine with 64GB RAM and with 15-minutes time out.  For each execution, we recorded the execution time and any output reports, error traces, and stack traces.  We then examined the output to determine the verdict and its veracity pertaining to the vulnerability.

\subsubsection{Ensuring Fairness}
\label{sec:vuln-tools-ensuring-fairness}

\paragraph{Consider only supported versions of Android:} \fixdroid was not evaluated on secure apps in Ghera because Android Studio version 3.0.1 was required to build the secure apps in Ghera and \fixdroid was available as a plugin only to Android Studio version 2.3.3.  Further, since we added benchmarks C4, I13, I15, and S4  after Ghera migrated to Android Studio version 3.0.1, \fixdroid was not evaluated on these benchmarks; hence, we evaluated \fixdroid on only 38 Ghera benchmarks.

\paragraph{Provide inputs as required by tools:} \covert \citep{Bagheri:TSE2015} and \dialdroid \citep{Bosu:2017} detect vulnerabilities stemming from inter-app communications, \eg collusion, compositional vulnerabilities.  So, we applied each of these tools in its default configuration to 33 Ghera benchmarks that included malicious apps.  For each benchmark, we provided the malicious app as input together with the benign app and the secure app.

\paragraph{Consider multiple operational modes:} \jaads \citep{Jaads:URL} operates in two modes: fast mode in which only intra-procedural analysis is performed ($\jaads_H$) and full mode in which both intra- and inter-procedural analyses are performed ($\jaads_E$).  Since the modes can be selected easily, we evaluated \jaads in both modes.

\qark \citep{Qark:URL} can analyze the source code and the APK of an app.  It decompiles the DEX bytecodes in an APK into source form.  Since the structure of reverse engineered source code may differ from the original source code and we did not know if this could affect the accuracy of \qark's verdicts, we evaluated \qark with both APKs ($\qark_A$) and source code ($\qark_S$).

\paragraph{Consider only supported API levels:}  Since the inception and evolution of tools are independent of the evolution of Android API levels, a tool may not support an API level (\eg the API level is released after the tool was last developed/updated) and, hence, it may fail to detect vulnerabilities stemming from APIs introduced in such unsupported API levels.  To control for this effect, we identified the API levels supported by tools.

\emph{Of the 14 tools, only three tools provide some information about the supported API levels (Android platform versions).}  Specifically, \jaads documentation states that \jaads will work for all Android platforms (as long as the platform version is provided as input).  \amandroid was successfully used in \citep{Wei:TOPS18} to process ICC-Bench benchmarks \citep{ICCBench:URL} that target API level 25; hence, we inferred \amandroid supports API level 25 and below.  \dialdroid tool repository contains versions of Android platform corresponding to API levels 3 thru 25 that are to be provided as input to the tool; hence, we inferred \dialdroid supports API levels 19 thru 25.

In the absence of such information, we conservatively assumed the tools supported API levels 19 thru 25, and this assumption is fair because 1) API level 19 and 25 were released in October 2013 and December 2016, respectively, (see \emph{API Levels} column in \Fref{tab:vul-time-distribution}), 2) all tools were last updated in 2013 or later (see \emph{Updated} column in \Fref{tab:vul-tools-info}), and 3) every Ghera benchmark APK used in the evaluation were built to run on every API level 19 thru 25.

\begin{table*}
  \centering
  \ifdef{\TopCaption}{
  \caption{Cumulative number of considered vulnerabilities discovered until a specific year (inclusive).  \emph{Tools} column is cumulative number of evaluated tools published until a specific year (inclusive). \emph{API Levels} column lists the API levels released in a specific year.}
  }{}
  \begin{tabular}{@{}lcccccccccr@{}}
    \toprule
    & \multicolumn{8}{c}{Cumulative \# of Considered Vulnerabilities} & & \\
    \cmidrule(l){2-9}
    Year & Crypto & ICC & Net & Perm & Store & Sys & Web & Total & \# Tools\footnotemark  & API Levels\\
    \midrule
    2011 & 0 & 7  & 0 & 0 & 2 & 0 & 0 & 9  & 0 & \\
    2012 & 0 & 7  & 0 & 0 & 2 & 0 & 4 & 13 & 0 & \\
    2013 & 4 & 7  & 0 & 0 & 2 & 0 & 7 & 20 & 1 & 19 \\
    2014 & 4 & 11 & 0 & 1 & 6 & 4 & 8 & 34 & 1 & 21 \\
    2015 & 4 & 14 & 1 & 1 & 6 & 4 & 9 & 39 & 3 & 22, 23 \\
    2016 & 4 & 14 & 1 & 1 & 6 & 4 & 9 & 39 & 4 & 24, 25 \\
    2017 & 4 & 14 & 2 & 1 & 6 & 4 & 9 & 40 & 9 & \\
    2018 & 4 & 16 & 2 & 1 & 6 & 4 & 9 & 42 & 13& \\
    \bottomrule
  \end{tabular}
  \ifundef{\TopCaption}{
  \caption{Cumulative number of considered vulnerabilities discovered until a specific year (inclusive).  \emph{Tools} column is cumulative number of evaluated tools published until a specific year (inclusive). \emph{API Levels} column lists the API levels released in a specific year.}
  }{}
  \label{tab:vul-time-distribution}
\end{table*}
\footnotetext[11]{In 2018, the total number of tools is 13 instead of 14 as the last updated date for \appcritique tool is unknown.}

\paragraph{Consider only applicable categories of vulnerabilities:} While each tool could be evaluated against all of 42 considered known vulnerabilities such an evaluation would be unfair as all tools are not geared to detect all kinds of vulnerabilities, \eg cryptography related vulnerability vs. storage related vulnerability.  Further, during app development, tools are often selected based on their capabilities pertaining to platform features/APIs used in the app.

To control for this effect, for each tool, we identified the categories of vulnerabilities that it was applicable to and evaluated it against only the vulnerabilities from these categories.  Even within categories, we ignored vulnerabilities if a tool was definitive not designed to detect them, \eg \mallodroid tool was evaluated only against SSL/TLS related vulnerabilities from Web category as the tool focuses on detecting SSL/TLS related vulnerabilities; see entry in \emph{Web} column for \mallodroid in \Fref{tab:vul-evals}.  For each tool, \Fref{tab:vul-tools-applicability} reports the applicable Ghera benchmark categories.

\begin{table*}
  \centering
  \ifdef{\TopCaption}{
    \caption{Applicability of vulnerability detection tools to various benchmark categories in Ghera.  ``\cmark'' denotes the tool is applicable to the vulnerability category in Ghera.  Empty cell denotes non-applicable cases.  ``*'' identifies non-academic tools.}
  }{}
  \begin{tabular}{@{}lccccccc@{}}
    \toprule
    & \multicolumn{7}{c}{API (Vulnerability) Categories} \\
    \cmidrule(l){2-8}
    Tool & Crypto & ICC & Net & Perm & Store & Sys & Web \\
    \midrule
    \amandroid & \cmark & \cmark & \cmark & & \cmark & & \cmark \\

    \androbugs* & \cmark & \cmark & \cmark & \cmark & \cmark & \cmark & \cmark \\

    \appcritique* & \cmark & \cmark & \cmark & \cmark & \cmark & \cmark & \cmark \\

    \covert & \cmark & \cmark & \cmark & \cmark & \cmark & \cmark & \cmark \\

    \devknox* & \cmark & \cmark & \cmark & \cmark & \cmark & \cmark & \cmark \\

    \dialdroid & \cmark & \cmark & \cmark & \cmark & \cmark & \cmark &  \cmark \\

    \fixdroid & \cmark & \cmark & \cmark & \cmark & \cmark & \cmark & \cmark \\

    \flowdroid & & \cmark & \cmark & & \cmark & & \\

    \horndroid & \cmark & \cmark & \cmark & & \cmark & & \cmark \\

    \jaads* & \cmark & \cmark & \cmark & \cmark & \cmark & \cmark & \cmark \\

    \mallodroid & & & & & & & \cmark \\

    \marvin* & \cmark & \cmark & \cmark & \cmark & \cmark & \cmark & \cmark \\

    \mobsf* & \cmark & \cmark & \cmark & \cmark & \cmark & \cmark & \cmark \\

    \qark* & \cmark & \cmark & \cmark & \cmark & \cmark & \cmark & \cmark \\
    \bottomrule
  \end{tabular}
  \ifundef{\TopCaption}{
    \caption{Applicability of vulnerability detection tools to various benchmark categories in Ghera.  ``\cmark'' denotes the tool is applicable to the vulnerability category in Ghera.  Empty cell denotes non-applicable cases.  ``*'' identifies non-academic tools.}
  }{}
  \label{tab:vul-tools-applicability}
\end{table*}

\paragraph{Consider the existence of vulnerabilities:} Expecting a tool to detect vulnerabilities that did not exist when the tool was developed/updated would be unfair.  In terms of the purpose of this evaluation, this is not a concern as the evaluation is less focused on individual tools and more focused on assessing the effectiveness of the existing set of vulnerability detection tools against considered known vulnerabilities.  In terms of the execution of this evaluation, this is not a concern as almost all of the considered vulnerabilities (39 out of 42) were discovered before 2016 (see \emph{Total} column in \Fref{tab:vul-time-distribution}) and almost all of the evaluated tools (at least 10 out of 14) were updated in or after 2016 (see \emph{\# Tools} column in \Fref{tab:vul-time-distribution}).

\subsection{Observations and Open Questions}
\label{sec:vul-tools-observations}

This section lists interesting observations along with interesting open questions that were uncovered but not answered in this evaluation.  While the data from this experiment supports the validity of the questions, the answers to these questions require more experimentation.
\subsubsection{Tools Selection}
\label{sec:vul-tools-obs-sel}

\paragraph{Open Questions 1 \& 2} Of the considered 64 solutions, 17 tools (including \amandroid) were intended to enable security analysis of Android apps; see \Fref{sec:vul-shallow-selection}.  In other words, 26.5\% of security analysis tools considered in this evaluation enable security analysis of Android apps.  Further, we have found these tools be useful in our research workflow, \eg Drozer \citep{Drozer:URL}, MobSF \citep{MobSF:URL}.  Hence, we believe that studying these tools may be useful.  Specifically, exploring two mutually related questions: \emph{1) how expressive, effective, and easy-to-use are tools that enable security analysis?} and \emph{2) are Android app developers and security analysts willing to invest effort in such tools?} may help both tool users and tool developers.

\paragraph{Observation 1} We rejected 39\% of tools (9 out of 23) considered in deep selection; see \Fref{sec:vul-deep-selection}.  Even considering the number of instances where the evaluated tools failed to process certain benchmarks (see numbers in square brackets in \Fref{tab:vul-evals}), such a low rejection rate is rather impressive and suggests \emph{tool developers are putting in effort to release robust security analysis tools}.  This number can be further improved by distributing executables (where applicable), providing complete and accurate build instructions (\eg versions of required dependencies) for local tools, providing complete and accurate information about execution environment (\eg versions of target Android platforms), and publishing estimated turn around times for remote tools.

\paragraph{Observation 2} If the sample of tools included in this evaluation is representative of the population of Android app security analysis tools, then \emph{almost every vulnerability detection tool for Android apps relies on static analysis, \ie 13 out of 14}; see \emph{S/D} column in \Fref{tab:vul-tools-info}.

\paragraph{Observation 3} Every vulnerability detection tool publicly discloses the category of vulnerabilities it tries to detect.  Also, almost all vulnerability detection tools are available as locally executable tools, \ie 13 out of 14; see \emph{L/R} column in \Fref{tab:vul-tools-info}.   So,  \emph{vulnerability detection tools are open with regards to their vulnerability detection capabilities}.  The likely reason is to inform app developers how the security of apps improves by using a vulnerability detection tool and encourage the use of appropriate vulnerability detection tools.

\paragraph{Observation 4} Ignoring tools with unknown update dates (``?'' in column 3 of \Fref{tab:vul-tools-info}) and considering that we conducted the evaluation between June 2017 and May 2018, 9 out of 13 tools are less than 1.5 years old (2017 or later) and 12 out of 13 are less than or about 3 years old (2015 or later).  Hence, the selected tools can be considered as current.  Consequently, \emph{the resulting observations are highly likely to be representative of the current state of the freely available Android app security analysis tools.}

\subsubsection{Vulnerability Detection Tools}
\label{sec:vul-tools-obs-vul}

\setlength{\tabcolsep}{\dimexpr\tabcolsep-1pt}
\begin{customTable}
  \centering
  \ifdef{\TopCaption}{
    \caption{Results from evaluating vulnerability detection tools. The number of benchmarks in each category is mentioned in parentheses.  In each category, empty cell denotes the tool is inapplicable to any of the benchmarks, N denotes the tool flagged both benign and secure apps as not vulnerable in every benchmark, X denotes the tool failed to process any of the benchmarks, and D denotes the tool flagged both benign and secure apps as vulnerable in every benchmark.  H/I/J/K denotes the tool was inapplicable to H benchmarks, flagged benign app as vulnerable and secure app as not vulnerable in I benchmarks, flagged both benign and secure app as not vulnerable in J benchmarks, and reported non-existent vulnerabilities in benign or secure apps in K benchmarks.  Along with D and N, I and J (in bold) contribute to TP and FN, respectively.  The number of benchmarks that a tool failed to process is mentioned in square brackets.  The number of benchmarks in which both benign and secure apps were flagged as vulnerable is mentioned in curly braces.  ``-'' denotes not applicable cases.  ``*'' identifies non-academic tools. ``?'' denotes unknown number.}
  }{}
  \begin{tabular}{@{}lccccccccccccc@{}}
  \toprule
  Tool & Crypto & ICC & Net & Perm & Store & Sys & Web & \multicolumn{2}{c}{\# Benchmarks} & \multicolumn{2}{c}{Benign} & Secure & Other \\
  \cmidrule(r){11-12}
   & (4) & (16) & (2) & (1) & (6) & (4) & (9) & Applicable & Known & TP & FN & TN & \\
  \midrule
  $\mbox{\amandroid}_1$ & & 7/\textbf{0}/\textbf{9}/3 & 1/\textbf{0}/\textbf{1}/0 & & 4/\textbf{0}/\textbf{1}/0 \{1\} & & 6/\textbf{0}/\textbf{3}/0 & 15 & 13 & 1 & 14 & 14 & 3\\
  $\mbox{\amandroid}_2$ & & X & &  & 1/\textbf{0}/\textbf{2}/0 [3] &  & 5/\textbf{0}/\textbf{0}/0 [4] & 25 & 23 & 0 & 3 & 3 & 0\\
  $\mbox{\amandroid}_3$ & & 8/\textbf{0}/\textbf{8}/2 & 1/\textbf{0}/\textbf{1}/0 & & 4/\textbf{0}/\textbf{2}/0 & & 6/\textbf{0}/\textbf{3}/0 & 14 & 12 & 0 & 14 & 14 & 2\\
  $\mbox{\amandroid}_4$ & & 13/\textbf{0}/\textbf{3}/2 & & & & & & 3 & 3 & 0 & 3 & 3 & 2\\
  $\mbox{\amandroid}_6$ & & & & & & & 6/\textbf{0}/\textbf{3}/0 & 3 & 3 & 0 & 3 & 3 & 0\\
  $\mbox{\amandroid}_7$ & 0/\textbf{2}/\textbf{2}/0 & & & & & & & 4 & 4 & 2 & 2 & 4 & 0\\
  \androbugs* & N & 0/\textbf{2}/\textbf{14}/3 & N & 0/\textbf{1}/\textbf{0}/0 & N & 0/\textbf{4}/\textbf{0}/0 & 0/\textbf{4}/\textbf{5}/1 & 42 & 39 & 11 & 31 & 42 & 4\\
  \appcritique* & 0/\textbf{2}/\textbf{2}/0 & N & N & N & 0/\textbf{3}/\textbf{3}/0 & N & 0/\textbf{2}/\textbf{7}/0 & 42 & ? & 7 & 35 & 42 & 0\\
  \covert & N & N & 1/\textbf{0}/\textbf{1}/0 & N & N & N & 8/\textbf{0}/\textbf{1}/0 & 33 & 30 & 0 & 33 & 33 & 0\\
  \devknox* & 0/\textbf{1}/\textbf{3}/0 & N & N & N & N & D & N & 42 & 40 & 5 & 37 & 38 & 0\\
  \dialdroid & N & N & 1/\textbf{0}/\textbf{1}/0 & N & N & N & 8/\textbf{0}/\textbf{1}/0 & 33 & 33 & 0 & 33 & 33 & 0\\
  \fixdroid & 0/\textbf{1}/\textbf{2}/2 & 13/\textbf{1}/\textbf{0}/0 & N & N & 0/\textbf{1}/\textbf{4}/0 & 0/\textbf{4}/\textbf{0}/0 & 0/\textbf{2}/\textbf{7}/0 & 25 & 25 & 9 & 16 & - & 2\\
  \flowdroid & & N & N & & N & & & 24 & 24 & 0 & 24 & 24 & 0\\
  \horndroid & N & 0/\textbf{1}/\textbf{15}/7 & N & & 0/\textbf{0}/\textbf{6}/1 & & 0/\textbf{0}/\textbf{9}/1 & 37 & 37 & 1 & 36 & 37 & 9\\
  $\mbox{\jaads}_H$* & N & 0/\textbf{2}/\textbf{14}/0 & N & N & N & N & 0/\textbf{4}/\textbf{5}/1 & 42 & 40 & 6 & 36 & 42 & 1\\
  $\mbox{\jaads}_E$* & N & 0/\textbf{2}/\textbf{14}/0 & N & N & N & N & 0/\textbf{4}/\textbf{5}/1 & 42 & 40 & 6 & 36 & 42 & 1\\
  \mallodroid & & & X & & & & 5/\textbf{0}/\textbf{1}/0 [3] & 4 & 4 & 0 & 1 & 1 & 0\\
  \marvin* & 0/\textbf{1}/\textbf{3}/0 & 0/\textbf{5}/\textbf{11}/3 & N & 0/\textbf{1}/\textbf{0}/0 & 0/\textbf{0}/\textbf{6}/2 & 0/\textbf{4}/\textbf{0}/0 & 0/\textbf{4}/\textbf{5}/0 & 42 & 39 & 15 & 27 & 42 & 5\\
  \mobsf* & 0/\textbf{1}/\textbf{3}/0 & 0/\textbf{5}/\textbf{11}/0 & N & 0/\textbf{1}/\textbf{0}/1 & 0/\textbf{1}/\textbf{5}/0 & 0/\textbf{4}/\textbf{0}/0 & 0/\textbf{3}/\textbf{6}/0 & 42 & 42 & 15 & 27 & 42 & 1\\
  $\mbox{\qark}_A$* & N & 0/\textbf{3}/\textbf{13}/0 & N & 0/\textbf{1}/\textbf{0}/0 & N & 0/\textbf{4}/\textbf{0}/0 & 0/\textbf{2}/\textbf{7}/0 & 42 & 40 & 10 & 32 & 42 & 0\\
  $\mbox{\qark}_S$* & N & 0/\textbf{3}/\textbf{13}/0 & N & 0/\textbf{1}/\textbf{0}/0 & N & 0/\textbf{4}/\textbf{0}/0 & 0/\textbf{2}/\textbf{7}/0 & 42 & 40 & 10 & 32 & 42 & 0\\
  \midrule
  \# Undetected & 1 & 5 & 2 & 0 & 2 & 0 & 2 & & & &\\
  \bottomrule
  \end{tabular}
  \ifundef{\TopCaption}{
    \caption{Results from evaluating vulnerability detection tools. The number of benchmarks in each category is mentioned in parentheses.  In each category, empty cell denotes the tool is inapplicable to any of the benchmarks, N denotes the tool flagged both benign and secure apps as not vulnerable in every benchmark, X denotes the tool failed to process any of the benchmarks, and D denotes the tool flagged both benign and secure apps as vulnerable in every benchmark.  H/I/J/K denotes the tool was inapplicable to H benchmarks, flagged benign app as vulnerable and secure app as not vulnerable in I benchmarks, flagged both benign and secure app as not vulnerable in J benchmarks, and reported non-existent vulnerabilities in benign or secure apps in K benchmarks.  Along with D and N, I and J (in bold) contribute to TP and FN, respectively.  The number of benchmarks that a tool failed to process is mentioned in square brackets.  The number of benchmarks in which both benign and secure apps were flagged as vulnerable is mentioned in curly braces.  ``-'' denotes not applicable cases.  ``*'' identifies non-academic tools. ``?'' denotes unknown number.}
  }{}
  \label{tab:vul-evals}
\end{customTable}
\setlength{\tabcolsep}{\dimexpr\tabcolsep+1pt}

Table \ref{tab:vul-evals} summarizes the results from executing tools to evaluate their effectiveness in detecting different categories of vulnerabilities.  In the table, the number of vulnerabilities (benchmarks) that a tool was applicable to (and applied to) in this evaluation and the number of vulnerabilities that were known when a tool was developed or last updated before this evaluation is given by \emph{Applicable} and \emph{Known} columns, respectively.

Every Ghera benchmark is associated with exactly one unique vulnerability \textit{v}, and its benign app exhibits \textit{v} while its secure app does not exhibit \textit{v}.  So, for a tool, for each applicable benchmark, we classified the tool's verdict for the benign app as either true positive (\ie \textit{v} was detected in the benign app) or false negative (\ie \textit{v} was not detected in the benign app).  We classified the tool's verdict for the secure app as either true negative (\ie \textit{v} was not detected in a secure app) or false positive (\ie \textit{v} was detected in a secure app).  Columns \emph{TP, FN}, and \emph{TN} in \Fref{tab:vul-evals} report true positives, false negatives, and true negatives, respectively.  False positives are not reported in the table as none of the tools except \devknox (observe the D for \devknox under \textit{System} benchmarks in \Fref{tab:vul-evals}) and \textit{data leakage} extension of \amandroid (observe the \{1\} for $\text{\amandroid}_1$ under \textit{Storage} benchmarks in \Fref{tab:vul-evals}) provided false positive verdicts.  Reported verdicts do not include cases in which a tool failed to process apps.

\paragraph{Observation 5} \emph{Most of the tools (10 out of 14) were applicable to every Ghera benchmark}; see \emph{\# Applicable Benchmarks} column in \Fref{tab:vul-evals}.  Except for \mallodroid, the rest of the tools were applicable to 24 or more Ghera benchmarks.  This observation is also true of \amandroid if the results of its pre-packaged extensions are considered together.

\paragraph{Observation 6} Based on the classification of the verdicts, 4 out of 14 tools detected none of the vulnerabilities captured in Ghera (``0'' in the \emph{TP} column in \Fref{tab:vul-evals}) considering all extensions of \amandroid as one tool.  Even in case of tools that detected some of the vulnerabilities captured in Ghera, none of the tools individually detected more than 15 out of the 42 vulnerabilities; see the numbers in the \emph{TP} column and the number of \textsc{N}'s under various categories in \Fref{tab:vul-evals}.  This number suggests that \emph{in isolation, the current tools are very limited in their ability to detect known vulnerabilities captured in Ghera.}

\paragraph{Observation 7} For 11 out of 14 tools,\footnote{\androbugs, \marvin, and \mobsf were the exceptions.} the number of false negatives was greater than 70\% of the number of true negatives; see \emph{FN} and \emph{TN} columns in \Fref{tab:vul-evals}.\footnote{We considered all variations of a tool as one tool, \eg \jaads.  We did not count \fixdroid as we did not evaluate it on secure apps in Ghera.}  This proximity between the number of false negatives and the number of true negatives suggests two possibilities: \emph{most tools prefer to report only valid vulnerabilities (\ie be conservative)} and \emph{most tools can only detect specific manifestations of vulnerabilities}.  Both these possibilities limit the effectiveness of tools in assisting developers to build secure apps.

\paragraph{Observation 8} Tools make claims about their ability to detect specific vulnerabilities or class of vulnerabilities.  So, we examined such claims.  For example, while both \covert and \dialdroid claimed to detect vulnerabilities related to communicating apps, neither detected such vulnerabilities in any of the 33 Ghera benchmarks that contained a benign app and a malicious app.  Also, while \mallodroid focuses solely on SSL/TLS related vulnerabilities, it did not detect any of the SSL vulnerabilities present in Ghera benchmarks.  We observed similar failures with \fixdroid.  See numbers in \emph{\# Applicable Benchmarks} and \emph{TP} columns for \covert, \dialdroid, \fixdroid, and \mallodroid in \Fref{tab:vul-evals}.  These failures suggest that \emph{there is a gap between the claimed capabilities and the observed capabilities of tools that could lead to vulnerabilities in apps.}

\paragraph{Observation 9} Different tools use different kinds of analysis under the hood to perform security analysis.  Tools such as \qark, \marvin, and \androbugs rely on \emph{shallow analysis} (\eg searching for code smells/patterns) while tools such as \amandroid, \flowdroid, and \horndroid rely on \emph{deep analysis} (\eg data flow analysis); see \emph{H/E} column in \Fref{tab:vul-tools-info}.  Combining this information with the verdicts provided by the tools (see \emph{TP} and \emph{FN} columns in \Fref{tab:vul-evals}) provides the number of vulnerabilities (not) detected by shallow analysis and deep analysis across various categories as listed in \Fref{tab:analysis-type}.  From this data, we observe \emph{tools that rely on deep analysis report fewer true positives and more false negatives than tools that rely on shallow analysis.}  We also observe \emph{tools that relied on shallow analysis detected all of the vulnerabilities detected by tools that relied on deep analysis}.

\begin{table}
  \centering
  \ifdef{\TopCaption}{
    \caption{Breakdown of vulnerabilities detected by shallow and deep analyses, and not detected by either analyses.}
  }{}
  \begin{tabular}{@{}lcccccccc@{}}
    \toprule
    Detected By & Crypto & ICC & Net & Perm & Store & Sys & Web & Total \\
    \midrule
    Deep Analysis & 2 & 2 & 0 & 0 & 0 & 0 & 4 & 8\\
    Shallow Analysis & 3 & 11 & 0 & 1 & 4 & 4 & 7 & 30\\
    Neither & 1 & 5 & 2 & 0 & 2 & 0 & 2 & 12\\
    \midrule
    Total & 4 & 16 & 2 & 1 & 6 & 4 & 9 & 42\\
    \bottomrule
  \end{tabular}
  \ifundef{\TopCaption}{
    \caption{Breakdown of vulnerabilities detected by shallow and deep analyses, and not detected by either analyses.}
  }{}
  \label{tab:analysis-type}
\end{table}

Further, among the evaluated tools, most academic tools relied on deep analysis while most non-academic tools relied on shallow analysis; see \emph{H/E} columns in \Fref{tab:vul-tools-info} and tools marked with * in \Fref{tab:vul-evals}.

\paragraph{Open Questions 3 \& 4} A possible reason for the poor performance of deep analysis tools could be they often depend on extra information about the analyzed app (\eg a custom list of sources and sinks to be used in data flow analysis), and we did not provide such extra information in our evaluation.  However, \jaads was equally effective in both fast (intra-procedural analysis only) ($\jaads_H$) and full (both intra- and inter-procedural analyses) ($\jaads_E$) modes, \ie true positives, false negatives, and true negatives remained unchanged across modes.  Also, \fixdroid was more effective than other deep analysis tools even while operating within an IDE; it was the fifth best tool in terms of the number of true positives.  Clearly, in this evaluation, shallow analysis tools seem to outperform deep analysis tools.  This observation raises two related questions: \emph{3) are Android app security analysis tools that rely on deep analysis effective in detecting vulnerabilities in general?} and \emph{4) are the deep analysis techniques used in these tools well suited in general to detect vulnerabilities in Android apps?}  These questions are pertinent because Ghera benchmarks capture known vulnerabilities and the benchmarks are small/lean in complexity, features, and size (\ie less than 1000 lines of developer created Java and XML files), and yet deep analysis tools failed to detect the vulnerabilities in these benchmarks.

\paragraph{Observation 10} Switching the focus to vulnerabilities, every vulnerability captured by \textit{Permission} and \textit{System} benchmarks were detected by some tool.  However, no tool detected any of the two vulnerabilities captured by \textit{Networking} benchmarks. Further, no tool detected 12 of 42 known vulnerabilities captured in Ghera (false negatives); see \emph{\# Undetected} row in \Fref{tab:vul-evals}.  In other words, \emph{using all tools together is not sufficient to detect the known vulnerabilities captured in Ghera.}

\begin{table}
  \centering
  \ifdef{\TopCaption}{
    \caption{Number of undetected vulnerabilities discovered in a specific year. ``--'' denotes zero undetected vulnerabilities were discovered.}
  }{}
  \begin{tabular}{@{}lcccccc@{}}
    \toprule
    Year & Crypto & ICC & Net & Store & Web & Total \\
    \midrule
    2011 & --&  2& --& --& --& 2 \\
    2012 & --& --& --& --& --& -- \\
    2013 &  1& --& --& --&  1& 2 \\
    2014 & --&  1& --&  2&  1& 4 \\
    2015 & --& --&  1& --& --& 1 \\
    2016 & --& --& --& --& --& -- \\
    2017 & --& --&  1& --& --& 1 \\
    2018 & --&  2& --& --& --& 2 \\
    \bottomrule
  \end{tabular}
  \ifundef{\TopCaption}{
    \caption{Number of undetected vulnerabilities discovered in a specific year. ``--'' denotes zero undetected vulnerabilities were discovered.}
  }{}
  \label{tab:vul-undetected-distribution}
\end{table}

In line with the observation made in \Fref{sec:vul-tools-experiment} based on \Fref{tab:vul-time-distribution} -- most of the vulnerabilities captured in Ghera were discovered before 2016, most of the vulnerabilities (9 out of 12) not detected by any of the evaluated tools were discovered before 2016; see \Fref{tab:vul-undetected-distribution}.

\paragraph{Open Questions 5 \& 6}
Of the 42 vulnerabilities, 30 vulnerabilities were detected by 14 tools with no or minimal configuration, which is collectively impressive.  Further, two questions are worth exploring: \emph{5) with reasonable configuration effort, can the evaluated tools be configured to detect the undetected vulnerabilities?} and \emph{6) would the situation improve if vulnerability detection tools rejected during tools selection are also used to detect vulnerabilities?}

\paragraph{Observation 11} Of the 14 tools, 8 tools reported vulnerabilities that were not the focus of Ghera benchmarks; see \emph{Other} column in \Fref{tab:vul-evals}.  Upon manual examination of these benchmarks, we found none of the reported vulnerabilities in the benchmarks.  Hence, \emph{with regards to vulnerabilities not captured in Ghera benchmarks, tools exhibit a high false positive rate.}

\paragraph{Observation 12} To understand the above observations, we considered the relevant APIs and security-related APIs from Ghera representativeness experiment described in \Fref{sec:ghera-evaluation}.  We compared the sets of relevant APIs used in benign apps (601 APIs) and secure apps (602 APIs).  We found that 587 APIs were common to both sets while 14 and 15 APIs were unique to benign apps and secure apps, respectively.  When we compared security-related APIs used in benign apps (117 APIs) and secure apps (108 APIs), 108 were common to both sets, and nine were unique to benign apps.  These numbers suggest that the benign apps (real positives) and the secure apps (real negatives) are similar in terms of the APIs they use and different in terms of how they use APIs, \ie different arguments/flags, control flow context.  Therefore, \emph{tools should consider aspects beyond the presence of APIs to successfully identify the presence of vulnerabilities captured in Ghera.}

\paragraph{Observation 13} We partitioned the set of 601 relevant APIs and the set of 117 security-related APIs used in benign apps into three sets: 1) \emph{common APIs} that appear in both true positive benign apps (flagged as vulnerable) and false negative benign apps (flagged as secure), 2) \emph{TP-only APIs} that appear only in true positive benign apps, and 3) \emph{FN-only APIs} that appear only in false negative benign apps.  The sizes of these sets in order in each partition were 440, 108, and 53, and 60, 39, and 18, respectively.  For both relevant and security-related APIs, the ratio of the number of TP-only APIs and the number of FN-only APIs is similar to the ratio of the number of true positives and the number of false negatives, \ie $108/53 \approx 39/18 \approx 30/12$.  This relation suggests, \emph{to be more effective in detecting vulnerabilities captured in Ghera, tools should be extended to consider FN-only APIs.}  Since vulnerabilities will likely depend on the combination of FN-only APIs and common APIs, such extensions should also consider common APIs.

\paragraph{Observation 14} We compared the sets of relevant APIs used in the 12 false negative benign apps (flagged as secure) and all of the secure apps (real negatives).  While 491 APIs were common to both sets, only 2 APIs were unique to benign apps.  In the case of security-related APIs, 77 APIs were common while only 1 API was unique to secure apps.  Consequently, in terms of APIs, false negative benign apps are very similar to secure apps.
Consequently, \emph{tools need to be more discerning to correctly identify benign apps in Ghera as vulnerable (\ie reduce the number of false negatives) without incorrectly identifying secure apps as vulnerable (\ie increase the number of false positives).}\newline

Besides drawing observations from raw numbers, we also drew observations based on evaluation measures.

While precision and recall are commonly used evaluation measures, they are biased --- ``they ignore performance in correctly handling negative cases, they propagate the underlying marginal prevalences (real labels) and biases (predicted labels), and they fail to take account the chance level performance'' by \citet*{Powers:JMLT11}.   So, we used informedness and markedness, which are unbiased variants of recall and precision, respectively, as evaluation measures.

As defined by \citet*{Powers:JMLT11}, \emph{Informedness} quantifies how informed a predictor is for the specified condition, and specifies the probability that a prediction is informed in relation to the condition (versus chance).  \emph{Markedness} quantifies how marked a condition is for the specified predictor and specifies the probability that a condition is marked by the predictor (versus chance).  Quantitatively, informedness and markedness are defined as the difference between true positive rate and false positive rate (\ie $TP / (TP + FN) - FP / (FP + TN)$) and the difference between true positive accuracy and false negative accuracy, (\ie $TP / (TP + FP) - FN / (FN + TN)$), respectively.\footnote{TP, FP, FN, and TN denote the number of true positive, false positive, false negative, and true negative verdicts, respectively.}  When they are positive, the predictions are better than chance (random) predictions.  When these measures are zero, the predictions are no better than chance predictions.  When they are negative, the predictions are perverse and, hence, worse than chance predictions.

In this evaluation, while we use the above quantitative definitions, we interpret \emph{informedness as a measure of how informed (knowledgeable) is a tool about the presence and absence of vulnerabilities} (\ie will a vulnerable/secure app be detected as vulnerable/secure?) and \emph{markedness as a measure of the trustworthiness (truthfulness) of a tool's verdict about the presence and absence of vulnerabilities}, \ie is an app that is flagged (marked) as vulnerable/secure indeed vulnerable/secure?

\begin{table*}
  \centering
  \ifdef{\TopCaption}{
    \caption{Precision, Recall, Informedness, and Markedness scores from tools evaluation. ``--'' denotes measure was undefined.  ``*'' identifies non-academic tools.}
  }{}
  \begin{tabular}{lrrrr}
    \toprule
    Tool & Precision & Recall & Informedness & Markedness \\
    \midrule
    $\mbox{\amandroid}_1$ & 0.500 & 0.067 & 0.000 & 0.000\\
    $\mbox{\amandroid}_2$ & -- & 0.000 & 0.000 & --\\
    $\mbox{\amandroid}_3$ & -- & 0.000 & 0.000 & --\\
    $\mbox{\amandroid}_4$ & -- & 0.000 & 0.000 & --\\
    $\mbox{\amandroid}_6$ & -- & 0.000 & 0.000 & --\\
    $\mbox{\amandroid}_7$ & 1.000 & 1.000 & 1.000 & 1.000\\
    \androbugs* & 1.000 & 0.262 & 0.262 & 0.575\\
    \appcritique* & 1.000 & 0.167 & 0.167 & 0.545\\
    \covert& -- & 0.000 & 0.000 & --\\
    \dialdroid & -- & 0.000 & 0.000 & --\\
    \devknox* & 0.556 & 0.119 & 0.024 & 0.062\\
    \fixdroid & 1.000 & 0.231 & -- & 0.000\\
    \flowdroid & -- & 0.000 & 0.000 & --\\
    \horndroid & 1.000 & 0.027 & 0.027 & 0.507\\
    \jaads* & 1.000 & 0.143 & 0.143 & 0.538\\
    \mallodroid & -- & 0.000 & 0.000 & --\\
    \marvin* & 0.938 & 0.357 & 0.333 & 0.540\\
    \mobsf* & 1.000 & 0.310 & 0.310 & 0.592\\
    \qark* & 1.000 & 0.333 & 0.333 & 0.600\\
    \bottomrule
  \end{tabular}
  \ifundef{\TopCaption}{
    \caption{Precision, Recall, Informedness, and Markedness scores from tools evaluation. ``--'' denotes measure was undefined.  ``*'' identifies non-academic tools.}
  }{}
  \label{tab:vul-tools-eval-measures}
\end{table*}

\Fref{tab:vul-tools-eval-measures} reports the informedness and markedness for the evaluated tools.  It also reports precision and recall to help readers familiar with precision and recall but not with informedness and markedness.  It reports the measures for each \amandroid plugin separately as we applied each plugin separately to different sets of benchmarks.  It does not report measures for each variation of \jaads and \qark separately as the variations of each tool were applied to the same set of benchmarks and provided identical verdicts.  For tools that did not flag any app as vulnerable (positive), markedness was undefined.  Informedness was undefined for \fixdroid as we had not evaluated it on secure (negative) apps.

\paragraph{Observation 15} Out of 13 tools, 6 tools were better informed than an uninformed tool about the presence and absence of vulnerabilities, \ie $0.14 \leq \text{informedness} \leq 0.33$; see \emph{Informedness} column in \Fref{tab:vul-tools-eval-measures}.  These tools reported a relatively higher number of true positives and true negatives; see \emph{TP} and \emph{TN} columns in \Fref{tab:vul-evals}.  At the extremes, while $\text{\amandroid}_7$ plugin was fully informed (\ie $\text{informedness} = 1.0$), the remaining tools and \amandroid plugins were completely uninformed about the applicable vulnerabilities they were applicable to, \ie $\text{informedness} \approx 0$ as they did not report any true positives.  Thus, \emph{tools need to be much more informed about the presence and absence of vulnerabilities to be effective.}

\paragraph{Observation 16} Out of 14 tools, 8 tools provided verdicts that were more trustworthy than random verdicts, \ie $0.5 \leq \text{markedness}$; see \emph{Markedness} column in \Fref{tab:vul-tools-eval-measures}.  The verdicts from $\text{\amandroid}_7$ plugin could be fully trusted with regards to the applicable vulnerabilities (benchmarks), \ie $\text{markedness} = 1.0$.  The verdicts of $\text{\amandroid}_1$ were untrustworthy, \ie markedness = 0.  The verdicts of \fixdroid cannot be deemed untrustworthy based on markedness score because we did not evaluate \fixdroid on secure apps.  The remaining tools and \amandroid plugins did not flag any apps from any of the applicable benchmarks as vulnerable, \ie no true positive verdicts.  Unlike in the case of informedness, \emph{while some tools can be trusted with caution, tools need to improve the trustworthiness of their verdicts to be effective.}

Both the uninformedness and unmarkedness (lack of truthfulness of verdicts) of tools could be inherent to techniques underlying the tools or stem from the use of minimal configuration when exercising the tools.  So, while both possibilities should be explored, \emph{the ability of tools to detect known vulnerabilities should be evaluated with extensive configuration before starting to improve the underlying techniques.}

\paragraph{Observation 17} In line with observation 9, \emph{in terms of both informedness and markedness, shallow analysis tools fared better than deep analysis tools}; see \emph{H/E} column in \Fref{tab:vul-tools-info} and \emph{Informedness} and \emph{Markedness} columns in \Fref{tab:vul-tools-eval-measures}.  Also, non-academic tools fared better than academic tools; see tools marked with * in \Fref{tab:vul-tools-eval-measures}.  Similar to observation 9, these measures also reinforce the need to explore questions 3, 4, and 5.

\begin{figure*}
  \centering
  \includegraphics[width=\textwidth]{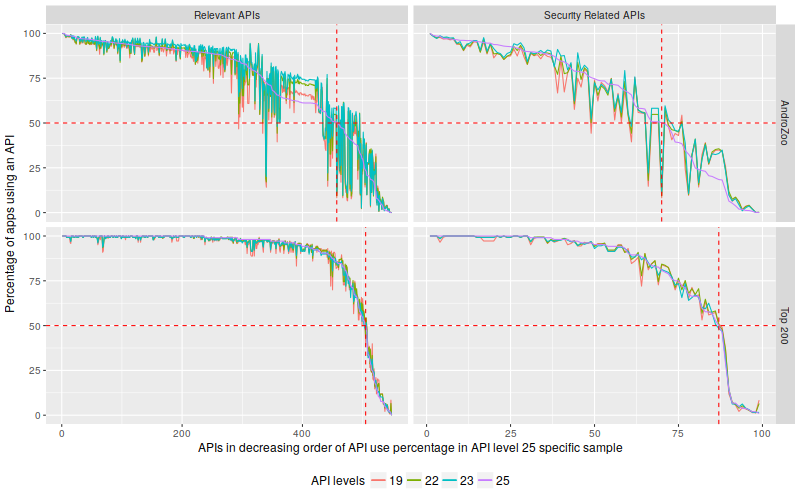}
  \caption{Percentage of apps that use APIs used in true positive benign Ghera apps.}
  \label{fig:detected-repr}
\end{figure*}

\begin{figure*}
  \centering
  \includegraphics[width=\textwidth]{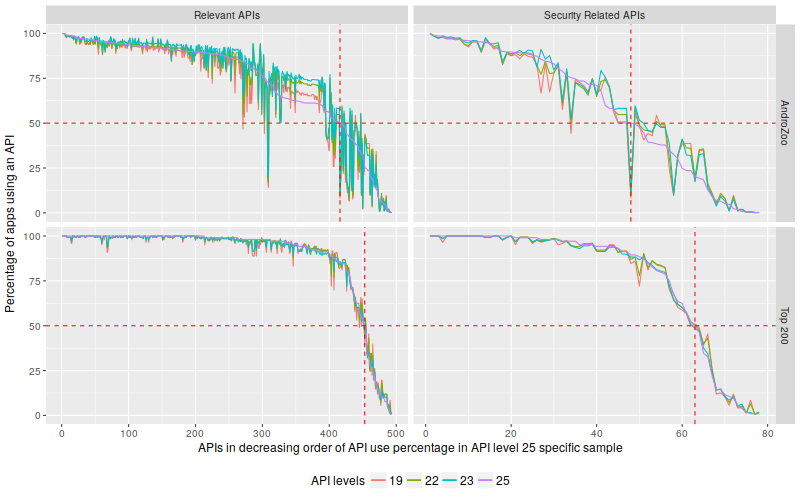}
  \caption{Percentage of apps that use APIs used in false negative benign Ghera apps.}
  \label{fig:undetected-repr}
\end{figure*}

\paragraph{Observation 18} We projected Ghera's representativeness information from \Fref{sec:ghera-evaluation} to APIs that were used in true positive benign apps (\ie common APIs plus TP-only APIs) and the APIs that were used in false negative benign apps.\footnote{We considered TP-only (FN-only) APIs along with common APIs as vulnerabilities may depend on the combination of TP-only (FN-only) APIs and common APIs.}  Barring the change in the upper limit of the x-axis, the API usage trends in figures \ref{fig:detected-repr} and \ref{fig:undetected-repr} are very similar to the API usage trend in \Fref{fig:representativeness}.  In terms of numbers, at least 83\% (457 out of 548) of relevant APIs and 70\% (70 out of 99) of security-related APIs used in true positive benign apps were used in at least 50\% of apps in both the Androzoo sample and the top-200 sample.  These numbers are 84\% (416 out of 493) and 61\% (48 out of 78) in case of false negative benign apps.  These numbers are very close to the API usage numbers observed when evaluating the representativeness of Ghera \Fref{sec:ghera-evaluation}.  Therefore, \emph{the effectiveness of tools in detecting vulnerabilities captured by Ghera benchmarks will extend to real-world apps.}

\paragraph{Observation 19} In terms of the format of the app analyzed by tools, \emph{all tools supported APK analysis.}  A possible explanation for this is that analyzing APKs helps tools cater to a broader audience: APK developers and APK users (\ie app stores and end users).

\paragraph{Observation 20} For tools that completed the analysis of apps (either normally or exceptionally), the median run time was 5 seconds with the lowest and highest run times being 2 and 63 seconds, respectively.  So, \emph{in terms of performance, tools that complete analysis exhibit good run times.}

\subsection{Threats to Validity}
\label{sec:vul-tools-TtV}

\paragraph{Internal Validity}  While we tried to execute the tools using all possible options, but with minimal or no extra configuration, we may not have used options or combinations of options that could result in more true positives and true negatives.  The same is true of extra configuration required by specific tools, \eg providing a custom list of sources and sink to \flowdroid.

Our personal preferences for IDEs (\eg Android Studio over Eclipse) and flavors of analysis (\eg static analysis over dynamic analysis) could have biased how we diagnosed issues encountered while building and setting up tools.  This preference could have affected the selection of tools and the reported set up times.

We have taken utmost care in using the tools, collecting their outputs, and analyzing their verdicts. However, our bias (\eg as creators of Ghera) along with commission errors (\eg incorrect tool use, overly permissive assumptions about API levels supported by tools) and omission errors (\eg missed data) could have affected the evaluation.

All of the above threats can be addressed by replicating this evaluation; specifically, by different experimenters with different biases and preferences than us and comparing their observations with our observations as documented in this manuscript and the artifacts repository (see \Fref{sec:artifacts}).

We used the categorization of vulnerabilities provided by Ghera as a coarse means to identify the vulnerabilities to evaluate each tool.  If vulnerabilities were mis-categorized in Ghera, then we may have evaluated a tool against an inapplicable vulnerability or not evaluated a tool against an applicable vulnerability.  This threat can be addressed by verifying Ghera's categorization of vulnerabilities or individually identifying the vulnerabilities that a tool is applicable to.\footnote{While evaluating the tools against all vulnerabilities seems like a solution to address this threat, we recommend against it as it would be unfair (as mentioned in \emph{Consider only applicable categories of vulnerabilities} paragraph in \Fref{sec:vuln-tools-ensuring-fairness}).}

\paragraph{External Validity} The above observations are based on the evaluation of 14 vulnerability detection tools.  While this is a reasonably large set of tools, it may not be representative of the population of vulnerability detection tools, \eg it does not include commercial tools, it does not include free tools that failed to build.  So, these observations should be considered as is only in the context of tools similar to the evaluated tools and more exploration should be conducted before generalizing the observations to other tools.

In a similar vein, the above observations are based on 42 known vulnerabilities that have been discovered by the community since 2011 and captured in Ghera since 2017.  While this is a reasonably large set of vulnerabilities and it covers different capabilities of Android apps/platforms, it may not be representative of the entire population of Android app vulnerabilities, \eg it does not include different manifestations of a vulnerability stemming from the use of similar yet distinct APIs, it does not include vulnerabilities stemming from the use of APIs (\eg graphics, UI) that are not part of the considered API categories, it does not include native vulnerabilities.  So, these observations can be considered as is only in the context of the considered 42 known vulnerabilities.  The observations can be considered with caution in the context of different manifestations of these vulnerabilities.  More exploration should be conducted before generalizing the observations to other vulnerabilities.

\subsection{Interaction with Tool Developers}
\label{sec:vul-tools-interaction-devs}

By the end of June 2017, we had completed a preliminary round of evaluation of \qark and \androbugs.  Since we were surprised by the low detection rate of both tools, we contacted both teams with our results and the corresponding responses were very different.

The \qark team acknowledged our study and responded to us with a new version of their tool after two months.  We evaluated this new version of \qark.  While the previous version of the tool flagged three benchmarks as vulnerable, the new version flagged ten benchmarks as vulnerable.  The evaluation reported in this manuscript uses this new version of \qark.

\androbugs team pointed out that the original version of their tool was not available on GitHub and, hence, we were not using the original version of \androbugs in our evaluation.  When we requested the original version of the tool, the team did not respond.\footnote{The corresponding GitHub ticket can be found at \url{https://github.com/AndroBugs/AndroBugs\_Framework/issues/16}.}

After these interactions, we decided not to communicate with tool developers until the evaluation was complete.  The rationale for this decision was to conduct a fair evaluation: evaluate tools as they are publicly available without being influenced by intermediate findings from this evaluation.

\subsection{Why did the evaluation take one year?}
\label{sec:vul-tools-evaluation-time}

\anonymize{In June 2017, we started gathering information about solutions related to Android app security.  By mid-June, we had performed both shallow and deep selection of tools.  At that time, Ghera repository had only 25 benchmarks, and we evaluated the selected tools on these benchmarks.

In July 2017, we started adding 17 new benchmarks to Ghera.  Also, we upgraded Ghera benchmarks to work with Android Studio 3.0.1 (from Android Studio 2.3.3).  So, in October 2017, we evaluated the tools against 25 revised benchmarks and 17 new benchmarks.

In November 2017, based on feedback at the PROMISE conference, we started adding secure apps to Ghera benchmarks along with automation support for functional testing of benchmarks.  This exercise forced us to change some of the existing benchmarks.  We completed this exercise in January 2018.

To ensure our findings would be fair and current, we re-evaluated the tools using the changed benchmarks, \ie a whole new set of secure apps along with few revamped benign and malicious apps.  We completed this evaluation in February 2018.

While the time needed for tools evaluation --- executing the tools and analyzing the verdicts --- was no more than two months, changes to Ghera and consequent repeated tool executions prolonged tools evaluation.

Between February and May 2018, we designed and performed the experiment to measure the representativeness of Ghera benchmarks.  Throttled remote network access to Androzoo, sequential downloading of apps from Androzoo, processing of 339K apps, and repetition of the experiment (to eliminate errors) prolonged this exercise.}

%% file: mal-tools-evaluation.tex
\section{Malicious Behavior Detection Tools Evaluation}
\label{sec:mal-tools-evaluation}

We decided to evaluate the effectiveness of malicious behavior detection tools against the malicious apps in Ghera benchmarks due to three reasons: 1) each Ghera benchmark contained a malicious app, 2) we had uncovered few malicious behavior detection tools during tool selection, and 3) we were in the midst of an elaborate tools evaluation.  Since the evaluation was opportunistic, incidental, exploratory, and limited, we did not expect to uncover report worthy results.  However, we did uncover strong results even in the limited scope; hence, we decided to report the evaluation.

\subsection{Tools Selection}
\label{sec:mal-tools-selection-mb}

For this evaluation, we considered the six tools that we rejected as malicious behavior detection tools during the shallow selection of tools for evaluating vulnerability detection tools (\Fref{sec:vul-tools-selection}):  \andrototal, \hickwall, \maldrolyzer, \nviso, StaDyna, and \virustotal.  Of these tools, we rejected StaDyna because it detected malicious behavior stemming from dynamic code loading and Ghera did not have benchmarks that used dynamic code loading \citep{Zhauniarovich:2015}.  Since the other five tools did not publicly disclose the malicious behaviors they can detect, we selected all of them for evaluation and considered them as applicable to all Ghera benchmarks.

Similar to \Fref{tab:vul-tools-info}, \Fref{tab:mal-tools-info} reports information about the malicious behavior detection tools selected for evaluation.

\setlength{\tabcolsep}{\dimexpr\tabcolsep-3pt}
\begin{table*}
  \centering
  \ifdef{\TopCaption}{
  \caption{Evaluated malicious behavior detection tools. \emph{S} and \emph{D} denote use of static analysis and dynamic analysis, respectively.  \emph{L} and \emph{R} denote the tool runs locally and remotely, respectively.  \emph{AV} stands for antivirus.  ``?'' denotes unknown information.}
  }{}
  \begin{tabular}{@{}lccccccr@{}} \toprule
    Tool & Commit Id /  & Updated & S/D & L/R & Uses AV & Set Up \\
    & Version & [Published] & & & Scanner & Time (sec)\\
    \midrule
    \andrototal \citep{Maggi:2013} & ? & ? [2013] & SD & R & Y & ? \\
    \hickwall \citep{Yuan:2014} & ? & ? [2014] & SD & R  & N & ? \\
    \maldrolyzer \citep{Maldrolyzer:URL} & 0919d46 & 2015 [2015] & S & L & N & 600 \\
    \nviso \citep{NViso:URL} & ? & ? [?] & SD & R & Y & ? \\
    \virustotal \citep{VirusTotal:URL} & ? & ? [?] & S & R & Y & ? \\
    \bottomrule
  \end{tabular}
  \ifundef{\TopCaption}{
  \caption{Evaluated malicious behavior detection tools. \emph{S} and \emph{D} denote use of static analysis and dynamic analysis, respectively.  \emph{L} and \emph{R} denote the tool runs locally and remotely, respectively.  \emph{AV} stands for antivirus.  ``?'' denotes unknown information.}
  }{}
  \label{tab:mal-tools-info}
\end{table*}
\setlength{\tabcolsep}{\dimexpr\tabcolsep+3pt}

\subsection{Experiment}
\label{sec:mal-tools-experiment}

We applied every selected tool in its default configuration to the \textit{malicious} app of every applicable Ghera benchmark.  Unlike in case of vulnerability detection tools, malicious behavior detection tools were evaluated on only 33 Ghera benchmarks because nine benchmarks --- 1 in Networking category and 8 in Web category --- did not have malicious apps as they relied on non-Android apps to mount man-in-the-middle exploits.

For this evaluation, we used the same execution set up from the evaluation of vulnerability detection tools.

\subsection{Observations}
\label{sec:mal-tools-observations}

\Fref{tab:mal-evals} reports the results from evaluating malicious behavior detection tools.  Like the results from evaluating vulnerability detection tools, these results are concerning but homogeneous.

\begin{table*}
  \centering
  \ifdef{\TopCaption}{
    \caption{Results from evaluating malicious behavior detection tools.  The number of benchmarks in each category is mentioned in parentheses.  X/Y denotes the tool deemed X benchmarks as PUPs and failed to flag Y benchmarks as malicious.}
  }{}
  \begin{tabular}{@{}lcccccccccc@{}}
    \toprule
    Tool & Crypto & ICC & Net & Perm & Sys & Store & Web & PUPs & FN\\
    & (4) & (16) & (1) & (1) & (4) & (6) & (1) & &\\
    \midrule
    \andrototal & 0/\textbf{4} & 0/\textbf{16} & 0/\textbf{1} & 0/\textbf{1} & 0/\textbf{4} & 0/\textbf{6} & 0/\textbf{1} & 0 & 33 \\
    \hickwall & 0/\textbf{4} & 0/\textbf{16} & 0/\textbf{1} & 0/\textbf{1} & 0/\textbf{4} & 0/\textbf{6} & 0/\textbf{1} & 0 & 33 \\
    \maldrolyzer & 0/\textbf{4} & 0/\textbf{16} & 0/\textbf{1} & 0/\textbf{1} & 0/\textbf{4} & 0/\textbf{6} & 0/\textbf{1} & 0 & 33 \\
    \nviso & 0/\textbf{4} & 9/\textbf{7} & 1/\textbf{0} & 1/\textbf{0} & 0/\textbf{4} & 4/\textbf{2} & 1/\textbf{0} & 16 & 17\\
    \virustotal & 4/\textbf{0} & 16/\textbf{0} & 1/\textbf{0} & 1/\textbf{0} & 0/4/\textbf{0} & 6/\textbf{0} & 1/\textbf{0} & 33 & 0\\
    \bottomrule
  \end{tabular}
  \ifundef{\TopCaption}{
    \caption{Results from evaluating malicious behavior detection tools.  The number of benchmarks in each category is mentioned in parentheses.  X/Y denotes the tool deemed X benchmarks as PUPs and failed to flag Y benchmarks as malicious.}
  }{}
  \label{tab:mal-evals}
\end{table*}

\paragraph{Observation 21} None of the evaluated malicious behavior detection tools publicly disclose the malicious behaviors they can detect.  Almost none of the malicious behavior detection tools are available as local services, \ie 4 out of 5; see \emph{L/R} column in \Fref{tab:mal-tools-info}.  So, \emph{malicious behavior detection tools are closed with regards to their detection capabilities}.  The likely reason is to thwart malicious behaviors by keeping malicious developers in the dark about current detection capabilities.

\paragraph{Observation 22} Recall that 33 benchmarks in Ghera are composed of two apps: a malicious app that exploits a benign app.  So, while malicious apps in Ghera are indeed malicious, 3 out of 5 tools did not detect malicious behavior in any of the malicious apps; see the \emph{FN} column in \Fref{tab:mal-evals}.  \virustotal flagged all of the malicious apps as \textit{potentially unwanted programs (PUPs)}, which is not the same as being malicious; it is more akin to flagging a malicious app as non-malicious.  \nviso flagged one half of the apps as PUPs and the other half of the apps as non-malicious.  In short, \emph{all malicious behavior detection tools failed to detect any of the malicious behaviors present in Ghera benchmarks}.

\paragraph{Observation 23} Since \andrototal, \nviso, and \virustotal rely on antivirus scanners to detect malicious behavior, the results suggest \emph{the malicious behaviors present in Ghera benchmarks will likely go undetected by antivirus scanners}; see \emph{Uses AV Scanners} column in \Fref{tab:mal-tools-info} and \emph{FN} column in \Fref{tab:mal-evals}.

\paragraph{Observation 24} Since \hickwall, \maldrolyzer, and \nviso rely on static and/or dynamic analysis to detect malicious behavior, the results suggest \emph{static and dynamic analyses used in these tools are ineffective in detecting malicious behaviors present in Ghera benchmarks}; see \emph{S/D} column in \Fref{tab:mal-tools-info} and \emph{FN} column in \Fref{tab:mal-evals}. \newline

\noindent Since we evaluated only five tools against 33 exploits, the above observations should not be generalized.  Nevertheless, \emph{the observations suggest 1) tools to detect malicious behavior in Android apps are likely to be ineffective and 2) evaluations that consider more malicious behaviors (exploits) and more tools are needed to accurately assess the effectiveness of existing malicious behavior detection tools and techniques.}

%% file: related-work.tex
\section{Related Work}
\label{sec:related-work}

Android security has generated considerable interest in the past few years.   This interest is evident by the sheer number of research efforts exploring Android security.   \citet{Sufatrio:2015} summarized such efforts by creating a taxonomy of existing techniques to secure the Android ecosystem.  They distilled the state of the art in Android security research and identified potential future research directions.  While their effort assessed existing techniques theoretically on the merit of existing reports and results, we evaluated existing tools empirically by executing them against a common set of benchmarks; hence, these efforts are complementary.

In 2016, \citet{Reaves:2016} systematized Android security research focused on application analysis by considering Android app analysis tools that were published in 17 top venues since 2010.  They also empirically evaluated the usability and applicability of the results of 7 Android app analysis tools.  In contrast, we evaluated 19 tools that detected vulnerabilities and malicious behaviors.  They used benchmarks from DroidBench \citep{DroidBench:URL}, six vulnerable real-world apps, and top 10 financial apps from Google Play store as inputs to tools.  While DroidBench benchmarks and vulnerable real-world apps were likely authentic (\ie they contained vulnerabilities), this was not the case with the financial apps.

In contrast, all of the 42 Ghera benchmarks used as inputs in our evaluation were authentic albeit synthetic.  While DroidBench focuses on ICC related vulnerabilities and use of taint analysis for vulnerability detection, Ghera is agnostic to the techniques underlying the tools and contains vulnerabilities related to ICC and other features of Android platform, \eg crypto, storage, web.  While their evaluation focused on the usability of tools (\ie how easy is it to use the tool in practice? and how well does it work in practice?), our evaluation focused more on the effectiveness of tools in detecting known vulnerabilities and malicious behavior and less on the usability of tools.  Despite these differences, the effort by Reaves \etal\ is closely related to our evaluation.

\citet{Sadeghi:TSE17} conducted an exhaustive literature review of the use of program analysis techniques to address issues related to Android security.  They identified trends, patterns, and gaps in existing literature along with the challenges and opportunities for future research.  In comparison, our evaluation also exposes gaps in existing tools.  However, it does so empirically while being agnostic to techniques underlying the tools, \ie not limited to program analysis.

More recently, \citet{Pauck:FSE18} conducted an empirical study to check if Android static taint analysis tools keep their promises. Their evaluation uses DroidBench, ICCBench \citep{ICCBench:URL}, and DIALDroidBench \citep{DIALDroid:URL} as inputs to tools.  Since the authenticity of the apps in these benchmark suites was unknown, they developed a tool to help them confirm the presence/absence of vulnerabilities in the apps and used it to create 211 authentic benchmarks.  Likewise, they created 26 authentic benchmarks based on the real-world apps from DIALDroiBench.  Finally, they empirically evaluated the effectiveness and scalability of 6 static taint analysis tools using these 237 benchmarks.  While their evaluation is very similar to our evaluation in terms of the goals --- understand the effectiveness of tools, there are non-trivial differences in the approaches and findings.  First, unlike their evaluation, our evaluation used Ghera benchmarks which are demonstrably authentic and did not require extra effort to ensure authenticity as part of the evaluation.  Further, while their evaluation is subject to bias and incompleteness due to manual identification of vulnerable/malicious information flows, our evaluation does not suffer from such aspects due to the intrinsic characteristics of Ghera benchmarks, \eg tool/technique agnostic, authentic.\footnote{Refer to Sections 2 and 3.4 of the work by \citet*{Mitra:PROMISE17} for a detailed description of characteristics of Ghera.}  Second, while they evaluated six security analysis tools, we evaluated 14 vulnerability detection tools (21 variations in total; see \Fref{tab:vul-evals}) (including 3 tools evaluated by Pauck \etal) along with five malicious behavior detection tools.  Further, while they evaluated only academic tools, we evaluated academic tools and non-academic tools.  Third, while their evaluation focused on tools based on static taint analysis, our evaluation was agnostic to the techniques underlying the tools.  Their evaluation was limited to ICC related vulnerabilities while our evaluation covered vulnerabilities related to ICC and other features of the Android platform, \eg crypto, storage, web.  Fourth, while their evaluation used more than 200 synthetic apps and 26 real-world apps, our evaluation used only 84 synthetic apps (\ie 42 vulnerable apps and 42 secure apps).  However, since each benchmark in Ghera embodies a unique vulnerability, our evaluation is based on 42 unique vulnerabilities.  In contrast, their evaluation is not based on unique vulnerabilities as not every DroidBench benchmark embodies a unique vulnerability \eg privacy leak due to constant index based array access vs. privacy leak due to calculated index based array access.
Finally, their findings are more favorable than our findings; even when limited to ICC related vulnerabilities.  Given the above differences in evaluations, the differences in findings are not surprising.

Given all of the above differences between these two evaluations that pursued very similar goals, we strongly urge researchers to consider these differences and associated experimental aspects while designing similar evaluations in the future.  Also, we believe a closer examination of existing empirical evaluations in software engineering is necessary to determine the pros and cons of the experimental setup used in these evaluations and identify the basic requirements of the experimental setup to create comparable and reproducible evaluations.

\citet{Zhou:SP2012} conducted a systematic study of the installation, activation, and payloads of 1260 malware samples collected from August 2010 thru 2011.  They characterized the behavior and evolution of malware.  In contrast, our evaluation is focused on the ability of tools to detect vulnerable and malicious behaviors.

%% file: artifacts.tex
\section{Evaluation Artifacts}
\label{sec:artifacts}

\begin{sloppypar}
Ghera benchmarks used in the evaluations described in this manuscript are available at \url{https://bitbucket.org/secure-it-i/android-app-vulnerability-benchmarks/src/RekhaEval-3}.

The code and input data used in the evaluation of representativeness of Ghera benchmarks are available in a publicly accessible repository: \url{https://bitbucket.org/secure-it-i/evaluate-representativeness/src/rekha-may2018-3}.  The respository also contains the output data from the evaluation and the instructions to repeat the evaluation.
\end{sloppypar}

A copy of specific versions of offline tools used in tools evaluation along with tool output from the evaluation are available in a publicly accessible repository: \url{https://bitbucket.org/secure-it-i/may2018}.  Specifically, \textit{vulevals} and \textit{secevals} folders in the repository contain artifacts from the evaluation of vulnerability detection tools using benign apps and secure apps from Ghera, respectively.  \textit{malevals} folder contains artifacts from the evaluation of malicious behavior detection tools using malicious apps from Ghera.  The repository also contains scripts used to automate the evaluation along with the instructions to repeat the evaluation.

To provide a high-level overview of the findings from tools evaluation, we have created a simple online dashboard of the findings at \url{https://secure-it-i.bitbucket.io/rekha/dashboard.html}.  The findings on the dashboard are linked to the artifacts produced by each tool in the evaluation.  We hope the dashboard will help app developers identify security tools that are well suited to check their apps for vulnerabilities and tool developers assess how well their tools fare against both known (regression) and new vulnerabilities and exploits.  We plan to update the dashboard with results from future iterations of this evaluation.

%% file: future-work.tex
\section{Future Work}
\label{sec:future-work}

Here are a few ways to extend this effort to help the Android developer community.

\begin{enumerate}
  \item Evaluate paid security analysis tools by partnering with tool vendors, \eg AppRay \citep{AppRay:URL}, IBM AppScan \citep{IBMAppScan:URL}, Klocwork \citep{Klocwork:URL}.
  \item Evaluate freely available Android security analysis tools that were considered but not evaluated in this tools evaluation, \eg ConDroid \citep{Anand:2012}, Sparta \citep{Ernst:CCS14}, StaDyna \citep{Zhauniarovich:2015}.
  \item Explore different modes/configurations of evaluated tools (\eg \amandroid) and evaluate their impact on the effectiveness of tools.
  \item Extend tools evaluation to consider new lean and fat benchmarks added to Ghera repository.  (Currently, Ghera contains 55 lean benchmarks: 14 new lean benchmarks and one deprecated lean benchmark.)
\end{enumerate}

%% file: summary.tex
\section{Summary}
\label{sec:summary}

When we started this evaluation, we expected many Android app security analysis tools to detect many of the known vulnerabilities.  The reasons for our expectation was 1) there has been an explosion of efforts in recent years to develop security analysis tools and techniques for Android apps and 2) almost all of the considered vulnerabilities were discovered and reported before most of the evaluated tools were last developed/updated.

Contrary to our expectation, the evaluation suggests that most of the tools and techniques can independently detect only a small number of considered vulnerabilities.  Even pooling all tools together, several vulnerabilities still remained undetected.  Also, we made several interesting observations such as tools using shallow analysis perform better than tools using deep analysis.

These observations suggest if current and new security analysis tools and techniques are to help secure Android apps, then they need to be more effective in detecting vulnerabilities; specifically, starting with known vulnerabilities as a large portion of real-world apps use APIs associated with these vulnerabilities.  A two-step approach to achieve this is 1) build and maintain an open, free, and public corpus of known Android app vulnerabilities in a verifiable and demonstrable form and 2) use the corpus to continuously and rigorously evaluate the effectiveness of Android app security analysis tools and techniques.

\section*{Acknowledgement}

We thank various readers and reviewers who read this manuscript and provided feedback to improve it.

%% file: catalog.tex
\section{Catalog of Considered Vulnerabilities}
\label{sec:catalog}

In this catalog, we briefly describe the 42 vulnerabilities captured in Ghera  (along with their canonical references) that were used in this evaluation.  Few vulnerabilities have generic references as they were discovered by Ghera authors while reading the security guidelines available as part of Android documentation \citep{AndroidSecTips:URL}.  Please refer the work by \citet*{Mitra:PROMISE17} for details about the repository and the initial set of vulnerabilities.

\subsection*{Acknowledgement} \anonymize{The authors would like to thank Aditya Narkar and Nasik Muhammad Nafi for their help in implementing 17 new benchmarks that are being cataloged as Ghera benchmarks for the first time in this paper and were used in the evaluations described in this paper.}

\subsection{Crypto}
\label{sec:cat-crypto}

Crypto APIs enable Android apps to encrypt and decrypt information and manage cryptographic keys.

\begin{enumerate}[{C}1]
  \item The result of encrypting a message twice using Block Cipher algorithm in \emph{ECB mode} is the message itself.  So, apps using Block Cipher algorithm in ECB mode (explicitly or due to default on Android platform) can  \emph{leak information} \citep{Egele:CCS13}.
  \item Encryption using Block Cipher algorithm in CBC mode with a \emph{constant Initialization Vector (IV)} can be broken by recovering the constant IV using plain text attack. So, apps using such encryption can \emph{leak information} \citep{Egele:CCS13}.
  \item Password-based encryption (PBE) uses a salt to generate a password-based encryption key.  If the salt is constant, the encryption key can be recovered with knowledge about the password.  Hence, apps \emph{using PBE with constant salt can leak information} \citep{Egele:CCS13}.
  \item Cipher APIs rely on unique keys to encrypt information.  If such keys are embedded in the app's code, then attackers can recover such keys from the app's code.  Hence, such apps are susceptible to both information leak and data injection \citep{Egele:CCS13}.
\end{enumerate}

\subsection{Inter Component Communication (ICC)}
\label{sec:cat-icc}

Android apps are composed of four basic kinds of components: 1) \textit{Activity} components display user interfaces, 2) \textit{Service} components perform background operations, 3) \textit{Broadcast Receiver} components receive event notifications and act on those notifications, and 4) \textit{Content Provider} components manage app data.  Communication between components in an app and in different apps (\eg to perform specific actions, share information) is facilitated via exchange of \textit{Intent}s.  Components specify their ability to process specific kinds of intents by using \emph{intent-filters}.

\begin{enumerate}[{I}1]
  \item Android apps can \emph{dynamically register broadcast receivers} at runtime.  Such receivers are automatically \emph{exported without any access restrictions} and, hence, can be accessed via ICC and exploited to perform unintended actions \citep{Chin:MOBISYS11}.
  \item A component can use a \emph{pending intent} to allow another component to act on its behalf.  When a pending intent is empty (\ie does not specify an action), it can be seeded (via interception) with an unintended action to be executed on behalf of the originating component \citep{CVE-2014-8609:URL}.
  \item To perform an action (\eg send email), users choose an activity/app from a list of \emph{activities ordered by priority}.  By using appropriate priorities, activities can gain unintended privilege over other activities \citep{Chin:MOBISYS11}.
  \item\label{list:i4} Implicit intents are processed by any \emph{qualifying service determined by intent filters} as opposed to a specific explicitly named service.  In case of multiple qualifying services, the service with the highest priority processes the intent.  By registering appropriate intent filters and by using appropriate priorities, services can gain unintended access to implicit intents \citep{Chin:MOBISYS11}.
  \item Pending intents can contain \emph{implicit intents}.  Hence, services processing implicit intents contained in pending intents are vulnerable like in I\ref*{list:i4} \citep{CVE-2014-8609:URL}.
  \item Apps can use \texttt{path-permissions} to control access to data exposed by content provider.  These permissions control access to a folder and not to its subfolders/descendants and their contents.  Incorrectly assuming the extent of these permissions can lead to information leak (read) and data injection (write) \citep{AndroidSecTips:URL}.
  \item\label{list:i7} Components that process \emph{implicit intents} are by default \emph{exported without any access restrictions}.  If such a component processes intents without verifying their authenticity (\eg source) and handles sensitive information or performs sensitive operations in response to implicit intents, then it can leak information or perform unintended actions \citep{Chin:MOBISYS11}.
  \item Broadcast receivers registered for \emph{system intents} (\eg low memory) from Android platform are by default \emph{exported without any access restrictions}.  If such a receiver services intents without verifying the authenticity of intents (\eg requested action), then it may be vulnerable like the components in I\ref*{list:i7} \citep{Chin:MOBISYS11}.
  \item Broadcast receivers respond to \emph{ordered broadcasts} in the order of priority.  By using appropriate priorities, receivers can modify such broadcasts to perform unintended actions \citep{Chin:MOBISYS11}.
  \item \emph{Sticky broadcast intents} are delivered to every registered receiver and saved in the system to be delivered to receivers that register in the future.  When such an intent is re-broadcasted with modification, it replaces the saved intent in the system, which can lead to information leak and data injection \citep{Chin:MOBISYS11}.
  \item Every activity is launched in a \emph{task}, a collection (stack) of activities.  An activity can declare its \emph{affinity} to be started in a specific task under certain conditions.  When the user navigates away from an activity X via the \emph{back} button, the activity below X on X's task is displayed.  This behavior along with task affinity and specific start order --- malicious activity started before benign activity --- can be used to mount a phishing attack \citep{Ren:USENIX15}.
  \item When an activity from a \emph{task in the background} (\ie none of its activities are being displayed) is resumed, the activity at the top of the task (and not the resumed activity) is displayed.  This behavior along with task affinity and specific start order --- malicious activity started after benign activity --- can be used to mount a phishing attack \citep{Ren:USENIX15}.
  \item When a \emph{launcher activity} is started, its task is created and it is added as the first activity of the task.  If the task of a launcher activity already exists with other activities in it, then its task is brought to the foreground, but the launcher activity is not started.  This behavior along with task affinity and specific start order --- malicious activity started before benign activity --- can be used to mount a phishing attack \citep{Ren:USENIX15}.
  \item In addition to declaring task affinity, an activity can request that it be moved to the affine task when the task is created or moved to the foreground (known as \emph{task reparenting}).  This behavior along with task affinity and specific start order --- malicious activity started before benign activity --- can be used to mount a denial-of-service or a phishing attack \citep{Ren:USENIX15}.
  \item Apps can request permission to \emph{perform privileged operations} (\eg send SMS) and offer interfaces (\eg broadcast receiver) thru which these privileged operations can be triggered.  Unless appropriately protected, these interfaces can be exploited to perform an operation without sufficient permissions \citep{Chin:MOBISYS11}.
  \item The \texttt{call} method of Content Provider API can be used to invoke any provider-defined method.  With reference to a content provider, this method can be invoked without any restriction leading to both information leak and data injection \citep{AndroidCP:URL}.
\end{enumerate}

\subsection{Networking}
\label{sec:cat-network}

Networking APIs allow Android apps to communicate over the network via multiple protocols.

\begin{enumerate}[{N}1]
  \item Apps can \emph{open server sockets} to listen to connections from clients; typically, remote servers.  If such sockets are not appropriately protected, then they can lead to \emph{information leak} and \emph{data injection} \citep{Jia:IEEE17}.
  \item Apps can communicate with remote servers via \emph{insecure TCP/IP connections}.  Such scenarios are susceptible to MitM attacks \citep{Jia:IEEE17}.
\end{enumerate}

\subsection{Permissions}
\label{sec:cat-perms}

In addition to system-defined permissions, Android apps can create and use custom permissions.  These permissions can be combined with
the available four protection levels (\ie normal, dangerous, signature, signatureOrSystem) to control access to various features and services.

\begin{enumerate}[{P}1]
  \item Permissions with normal protection level are automatically granted to requesting apps during installation.  Consequently, any component or its interface protected by such ``normal'' permissions will be accessible to every installed app \citep{CVE-2014-1977:URL}.
\end{enumerate}

\subsection{Storage}
\label{sec:cat-storage}

Android provides two basic options to store app data.
\begin{enumerate}
  \item \emph{Internal Storage} is best suited to store files private to apps.  Every time an app is uninstalled, its internal storage is purged.  Starting with Android 7.0 (API 24), files stored in internal storage cannot be shared with and accessed by other apps \citep{AndroidSecTips:URL}.
  \item \emph{External Storage} is best suited to store files that are to be shared with other apps or persisted even after an app is uninstalled.  While public directories are accessible by all apps, app-specific directories are accessible only by corresponding apps or other apps with appropriate permission \citep{AndroidSecTips:URL}.
\end{enumerate}

\begin{enumerate}[{S}1]
  \item\label{list:s1} \emph{Files stored in public directories on external storage} can be accessed by an app with appropriate permission to access external storage.  This aspect can be used to tamper data via data injection \citep{AndroidSecTips:URL}.
  \item The same aspect from S\ref*{list:s1} can lead to information leak \citep{AndroidSecTips:URL}.
  \item Apps can \emph{accept paths to files in the external storage from external sources and use them without sanitizing them}.  A well-crafted file path can be used to read, write, or execute files in the app's private directory on external storage (directory traversal attack) \citep{CVE-2014-5319:URL}.
  \item Apps can \emph{copy data from internal storage to external storage}.  As a result, information could be leaked if such apps accept input from untrusted sources to determine the data to be copied \citep{CVE-2014-1566:URL}.
  \item \texttt{SQLiteDatabase.rawQuery()} method can be used by apps to serve data queries.  If such uses rely on external inputs and use non-parameterized SQL queries, then they are susceptible to \emph{sql injection} attacks \citep{CVE-2014-8507:URL}.
  \item Content Provider API support \texttt{selectionArgs} parameter in various data access operations to separate selection criteria and selection parameters.  App that do not use this parameter are be susceptible to \emph{sql injection} attacks \citep{CVE-2014-8507:URL}.
\end{enumerate}

\subsection{System}
\label{sec:cat-system}
System APIs enable Android apps to access low-level features of the Android platform like process management, thread management, runtime permissions, etc.

Every Android app runs in its own process with a unique Process ID (PID) and a User ID (UID).  All components in an app run in the same process.  A permission can be granted to an app at installation time or at run time.  All components inherit the permissions granted to the containing app at installation time.  If a component in an app is protected by a permission, only components that have been granted this permission can communicate with the protected component.

\begin{enumerate}[{Y}1]
  \item\label{list:y1} During IPC, \texttt{checkCallingOrSelfPermission} method can be used to check if the calling process or the called process has permission P.  If a component with permission P uses this method to check if the calling component has permission P, then improper use of this method can \emph{leak privilege} when the calling component does not have permission P \citep{AndroidBinder:URL}.
  \item\label{list:y2} \texttt{checkPermission} method can be used to check if the given permission is allowed for the given PID and UID pair.  \texttt{getCallingPID} and \texttt{getCallingUID} methods of Binder API can be used to retrieve the PID and UID of the calling process.  In certain situations, they return PID/UID of the called process.  So, improper use of these methods by a called component with given permission can \emph{leak privilege} \citep{AndroidBinder:URL}.
  \item During IPC, \texttt{enforceCallingOrSelfPermission} method can be used to check if the calling process or the called process has permission P. Like in Y\ref*{list:y1}, improper use of this method can \emph{leak privilege} \citep{AndroidBinder:URL}.
  \item \texttt{enforcePermission} method can be used to check if the given permission is allowed for the given PID and UID pair.  Like in Y\ref*{list:y2}, improper use of this method along with \texttt{getCallingPID} and \texttt{getCallingUID} can \emph{leak privilege} \citep{AndroidBinder:URL}.
\end{enumerate}

\subsection{Web}
\label{sec:cat-web}

Web APIs allow Android apps to interact with web servers both with and without SSL/TLS, display web content through \texttt{WebView} widget, and control navigation between web pages via \texttt{WebViewClient} class.

\begin{enumerate}[{W}1]
  \item Apps connecting to remote servers via HTTP (as opposed to HTTPS) are susceptible to \emph{information theft} via \emph{Man-in-the-Middle (MitM)} attacks \citep{Tendulkar:E14}.
  \item\label{list:w1} Apps can employ \texttt{HostnameVerifier} interface to perform custom checks on hostname when using SSL/TLS for secure communication.  If these checks are incorrect, apps can end up connecting to malicious servers and be targets of malicious actions \citep{Tendulkar:E14}.
  \item In secure communication, apps employ \texttt{TrustManager} interface to check the validity and trustworthiness of presented certificates.  Like in W\ref*{list:w1}, if these checks are incorrect, apps can end up trusting certificates from malicious servers and be targets of malicious actions \citep{Tendulkar:E14}.
  \item Intents can be embedded in URIs.  Apps that do not handle such intents safely (\eg check intended app) can \emph{leak information} \citep{ChromeDev:URL}.
  \item Web pages can access information local to the device (\eg GPS location).  Apps that allow such access without explicit user permission can \emph{leak information} \citep{CVE-2014-0806:URL}.
  \item When \texttt{WebView} is used to display web content, JavaScript code executed as part of the web content is executed with the permissions of the host app.  Without proper checks, malicious JavaScript code can get access to the app's resources, e.g. private files \citep{Chin:SPRINGER13}.
  \item When loading content over a secure connection via \texttt{WebView}, host app is notified of SSL errors via \texttt{WebViewClient}.  Apps ignoring such errors can enable MitM attacks \citep{Tendulkar:E14}.
  \item When a web resource (\eg CSS file) is loaded in \texttt{WebView}, the load request can be validated in \texttt{shouldInterceptRequest} method of \texttt{WebViewClient}.  Apps failing to validate such requests can allow loading of malicious content \citep{Chin:SPRINGER13}.
  \item \sloppypar{When a web page is loaded into \texttt{WebView}, the load request can be validated in \texttt{shouldOverridUrlLoading} method of \texttt{WebViewClient}.}  Apps failing to validate such requests can allow loading of malicious content \citep{Chin:SPRINGER13}.
\end{enumerate}